\documentclass[a4paper,fleqn,usenatbib]{mnras}
\usepackage{amsmath}
\usepackage{txfonts}
\usepackage{caption}
\usepackage[T1]{fontenc}
\usepackage{ae,aecompl}
\hypersetup{colorlinks, citecolor=green, filecolor=black, linkcolor=blue, urlcolor=blue }

\usepackage{graphicx}
\usepackage{amssymb}
\usepackage{dblfloatfix}
\usepackage{breqn}
\usepackage{hyperref}
\hypersetup{
    colorlinks=true,
    linkcolor=blue,
    filecolor=magenta,      
    urlcolor=cyan,
}

\title[Reverberation in GX 339-4]{Reverberation Reveals the Truncated Disc in the Hard State of GX 339-4}

\author[Mahmoud, Done \& De Marco]{
Ra'ad D. Mahmoud$^{1}$\thanks{E-mail: ra'ad.d.mahmoud@durham.ac.uk}, 
Chris Done$^{1}$ \&
Barbara De Marco$^{2}$
\\
$^{1}$Department of Physics, University of Durham, South Road, Durham DH1 3LE, UK\\
$^{2}$Nicolaus Copernicus Astronomical Center, Polish Academy of Sciences, Bartycka 18, PL-00-716 Warsaw, Poland\\
}

\date{Accepted XXX. Received YYY; in original form ZZZ}

\pubyear{2019}

\hypersetup{draft}
\begin{document}
\label{firstpage}
\pagerange{\pageref{firstpage}--\pageref{lastpage}}
\maketitle

% Abstract of the paper
\begin{abstract}
The nature and geometry of the hard state in black hole binaries is controversial. The broadband continuum spectrum and fast variability properties can be explained in a model where the inner disc evaporates into a geometrically thick, hot flow. However these models are challenged by the persistent detection of an extremely broad iron line, which requires that the disc extends down to the last stable orbit of a high spin black hole. This line width can be considerably reduced if the Comptonisation continuum is multi-component rather than single temperature, but such models are highly degenerate. Here we show a specific model of a radially stratified continuum coupled to a model of propagating fluctuations, fit to some of the best hard state data from GX 339-4. This full spectral-timing model can fit the time averaged spectrum, the power spectra in different energy bands, and the frequency dependent lags between these bands. For the first time we also include disc reverberation and show that this same spectral-timing model successfully \textit{predicts} the lag-energy spectra on all timescales. This gives a more robust method to determine the inner radius of the disc, which is of order $20~R_g$, i.e. significantly truncated. This opens up the way to use the fast variability spectral-timing data to trace the source geometry of black hole binaries in all states.
\end{abstract}


%%%%%%%%%%%%%%%%%%%%%%%%%%%%%%%%%%%%%%%%%%%%%%%%%%

%%%%%%%%%%%%%%%%% BODY OF PAPER %%%%%%%%%%%%%%%%%%

\section{Introduction}
\label{sec:INTRODUCTION}
Black hole X-ray binaries (BHXRBs) show (at least) two distinct spectral states. The hard state is predominantly seen at low luminosities where the spectrum is dominated by Compton scattering from hot plasma at $\sim 100$~keV, while the soft state is seen at high luminosities, where the spectrum is dominated by a thermal component at $\sim 1$~keV (\citealt{MR06}). There is widespread consensus that the soft state is produced in a cool, geometrically thin, optically thick disc which extends down to the innermost stable circular orbit (ISCO; e.g. \citealt{GD04}; \citealt{S10}), this disc being described by Shakura \& Sunyaev (1973; see also \citealt{NT73} for the relativistic form). There is less consensus for the hard state. All of the observed spectral and timing properties, and their evolution with changing mass accretion rate, match very well with a model where the inner disc progressively evaporates into a hot, geometrically thick, and optically thin, radiatively-inefficient accretion flow as the source luminosity drops (RIAF; \citealt{YQN03}; see the review by \citealt{DGK07}, also \citealt{EMN97}; \citealt{ID11}, hereafter ID11; \citealt{ID12a}). The only exception is from observations of a broad iron line in the brightest hard state, where the inferred line width is so large as to require a disc extending close to the ISCO of a high spin black hole, completely inconsistent with a truncated disc. One of the original influential observations of this suffered from instrumental pileup (\citealt{M06}, as demonstrated by \citealt{DDT10}), however independent analysis on Rossi X-ray Timing Explorer (RXTE) proportional counter data data show the same features (\citealt{G15}). These spectra are more complex than can be fit with reflection by a truncated disc from a single Comptonisation component. However this additional complexity can be equally well modeled by either dramatically relativistically-smeared reflection (\citealt{F14}; \citealt{P16}), or by the inclusion of a second Comptonisation continuum (see for Cyg X-1 e.g. \citealt{M08}; \citealt{N11}; \citealt{B17}, and see for GX 339-4 e.g. \citealt{KDD14}; \citealt{F15}, and Appendix A in \citealt{BZ16}). Spectral analysis alone cannot distinguish between these (and other) possibilities (\citealt{N11}; \citealt{BZ16}; \citealt{DAM18}). Physical plausibility arguments are likewise not conclusive; on the one hand the inhomogeneous Compton models are not in conflict with the highly successful truncated disc geometry, but on the other hand, adding multiple Compton components can make spectral fitting alone highly degenerate. An independent measure of the truncation radius is therefore absolutely required to break this degeneracy between models, and the most compelling source for this additional information lies in the fast ($0.01-10$~s) timing properties of the hard state (\citealt{vdK89}; \citealt{VN97}; \citealt{WU09}; \citealt{AD18}).

The hard state exhibits extreme variability on short timescales, with fractional root-mean-square variability amplitudes of up to $40$~\% (\citealt{MMB11}). The processes driving this strong variability must be related to the physical properties of the source, and herein lies the information which could break the degeneracies on the source geometry. A wealth of data from fast-timing instruments shows that the fast variability properties themselves are highly energy-dependent, with these data routinely showing fluctuations in higher energy bands lagging fluctuations at low energy bands by an amount which decreases with the timescale of the fluctuation (\citealt{MK89}; \citealt{RGC99}; \citealt{WU09}; \citealt{U14}; \citealt{G14}). Compton scattering should imprint a lag as a function of energy due to the additional light travel time required for each successive scattering order, but this is much shorter than observed lags and has no dependence on fluctuation timescale (\citealt{N99}). Instead, these lags are now generally interpreted as being related to the propagation timescale of fluctuations through the accretion flow. In the simplest models, slow fluctuations are produced at large radii so have to propagate through the entire flow, with a long lag time, whereas faster fluctuations are produced at smaller radii, so have a shorter distance and hence shorter lag time (\citealt{L97}; \citealt{KCG01}; \citealt{AU06}). These lags can then be imprinted on the lightcurves if the spectral shape in the flow is also radius-dependent.

The truncated disc models have a clear spectral stratification, between the disc at large radii, and the hot flow at smaller radii. Lags naturally arise from propagation of mass accretion rate fluctuations from the thermal disc to the Comptonising flow (\citealt{AU06}; \citealt{U11}), with the delay coming from the propagation timescale (\citealt{R17a}). However this picture alone does not explain why relatively long lags are still seen when comparing two bands which have negligible disc contribution (i.e. both bands above $\sim 2-3$~keV). At their largest amplitude, these lags can reach upwards of $~0.1$~seconds (\citealt{N99}; \citealt{U11}). This can instead be explained if the shape of the Compton spectrum itself is dependent on radius \textit{within} the hot flow, providing an energetic marker by which energy-dependent lags can be resolved (\citealt{KCG01}). In this case, mass accretion rate fluctuations first excite Comptonisation in the softer, outer part of the hot flow, before propagating in to excite Comptonisation in the harder, inner region. Physically, we should expect exactly this spectral inhomogeneity with radius, given that the inner regions will be starved of seed photons from the disc, resulting in a different spectral shape in the inner flow compared to the outer flow (\citealt{PV14}). With a harder spectral shape in the inner region, higher energy bands should be dominated by emission from the inner regions of the hot flow, while lower (but still Compton dominated) bands will be dominated by the outer parts of the flow, and so the lag we observe is simply a diluted form of the physical propagation lag (\citealt{MD18a}, hereafter MD18a).

The energy- and fluctuation-dependent lags in the fast variability therefore strongly favour models with inhomogeneous Comptonisation in a radially extended hot flow, in agreement with results of reflection fits which allow for a truncated disc geometry. However, fitting the propagation lags does not give an unambiguous size scale for the truncated disc. The low frequency break in the power spectrum gives the timescale for the slowest fluctuations generated in the hot flow, but we do not know \textit{a priori} how the timescales on which fluctuations are generated locally and drift radially relate to radius. Even when assuming that these timescales both act on a local `viscous' timescale, the power spectrum and lags produced by a geometry with a large truncation radius and fast viscous timescale can be indistinguishable from one produced by a geometry with a small truncation radius and slow viscous timescale (MD18a).

ID11 derive the viscous timescale in the flow using the additional information from the low-frequency Quasi-Periodic Oscillation (QPO) often seen in the power spectra of BHXRBs. This most probably arises from Lense-Thirring precession of the entire hot flow (\citealt{F07}; \citealt{IDF09}), and this timescale is particularly sensitive to the outer radius of the hot flow. Empirically, the QPO tracks the low-frequency break in the power spectrum as the source geometry evolves as a function of mass accretion rate (\citealt{WvdK99}; \citealt{B05}). This gives a way to derive the viscous timescale as a function of radius, assuming that Lense-Thirring precession is indeed the origin of the QPO (\citealt{I16}).

Here we use X-ray reverberation to provide an independent measure of the inner disc size scale. A change in the X-ray flux illuminating the disc gives rise to a light travel time delay in the response of the reprocessed emission, which includes the reflected continuum, iron line, and the energy absorbed by the disc which is re-emitted as a thermal component. This mechanism is more commonly used in Active Galactic Nuclei (AGN) as unlike BHXRBs they have substantial variability on timescales close to the light travel time of a few $R_g$ (e.g. \citealt{WH91}; \citealt{K13}; \citealt{GD14}). However, more recently these soft lag features have been seen in the energy-dependent high-frequency variability of BHXRBs (\citealt{U11}; \citealt{DM15}; \citealt{DM16}; \citealt{DM17}; hereafter DM17). Interpreting the reverberation lags is straightforward, although it depends on the assumed spectral model (due to dilution, see \citealt{U14}), on the underlying propagation lags, and on the illumination geometry. Here we simultaneously address all these issues for the first time. We incorporate reverberation into our full spectral-timing propagation model, and apply this to one of the best available datasets for GX 339-4, observed with XMM-Newton and NuSTAR. There are multiple datasets showing the reverberation lag for this object, both in the brighter fast rise to outburst (\citealt{DM15}), and during the fainter slow decline phase (DM17). The power spectra during the fast rise typically show more complexity than during the slow decline (\citealt{R17a}; \citealt{R17b}, \citealt{MD18b}, hereafter MD18b), so we pick the brightest of the slow decline hard datasets in order to maximise signal-to-noise with minimal source complexity.

We fit the joint XMM-Newton and NuSTAR energy spectra with two Compton components in addition to their reflection from the disc and the intrinsic and reprocessed disc emission. We use these components to build the fast variability model of fluctuations propagating through the three spectral regions, with the viscous timescale of the truncated disc set by the QPO - low-frequency break relation. We find a best fit when the hot flow extends from $\sim 19 - 4~R_g$, with a transition from \textit{soft} to \textit{hard}Comptonisation at around $6~R_g$. We then predict the lag-energy spectra in three different frequency bands and compare these to the data. The match is good, especially when we include the response of the ionization state of the accretion disc to the changing illumination. This enhances the change in reflected continuum above and below the FeK$\alpha$ line, making it appear that the response at the line is suppressed.

The reverberation lag is consistent with an inner disc radius of $19.5~R_g$ in these hard state data. This is a factor $\sim 4$ smaller than the simple measure of light travel time found in DM17, as the finite width of the transfer function leads to most of the reverberation signal coming from further out on the disc. By contrast, our result is a factor $\sim 8$ larger than the inner radius derived from the iron line profile using only a single Comptonisation continuum (\citealt{WJ18}). We conclude that the combined spectral-timing data strongly support the truncated disc models for the hard state, and that fitting the energy dependent power spectra and lags strongly requires a continuum model where there is more than a single Comptonisation component. Our reverberation size scale estimate is compatible with that derived using propagation alone with the viscous timescale set by the QPO-low frequency break relation, showing yet more evidence that the QPO is indeed due to Lense-Thirring precession of the flow.

\section{Our Data}
\label{sec:Data}
For our spectral fits, we primarily make use of data from GX 339-4 in the slow decay phase of its 2015 outburst (ObsID: 0760646201, hereafter O1), following the reduction procedure described in DM17, with updated calibration files (as of May 2018). This data was gathered using the XMM-Newton EPIC-pn instrument in timing mode over a $14.9$~ks exposure. During this observation, the source exhibits root-mean-square variability amplitude of $0.28 \pm 0.01$, confirming its position on the slow decay phase of the outburst. We also extend these data to higher energies by using a simultaneous $21.6$~ks observation from NuSTAR. The NuSTAR count rates are generally too low for sensitive fast timing, but these data do extend the spectral range up to $70$~keV, which allows us to constrain the high-energy cutoff of the Comptonisation spectrum. This cutoff is important to constrain the bolometric luminosity of the hard Compton component, which in turn is crucial when considering the fraction of thermal emission which originates from reprocessing, as we will discuss in the next section.

The three lightcurves we use for timing model comparison are extracted from the Low ($0.3-0.7$~keV), Intermediate ($0.7-1.5$~keV) and High ($3-5$~keV) energy bands of the XMM O1 observation. These light curves are identical to the `very soft', `soft' and `hard' light curves in DM17, but we change the nomenclature here to avoid confusion with our spectral model components. For each lightcurve we calculate Poisson-noise subtracted power spectra and time lags by ensemble averaging over 298 intervals, each containing $2^{12}$ time bins of $0.012$~s length. This gives power spectra and time lags in the frequency range $0.02-41.6$~Hz denoted respectively $P_i(f)$ and $\tau_{ij}(f)$ where $i,j \in [L,\,I,\,H]$ are from our set of three energy bands.

\begin{figure}
	\includegraphics[width=\columnwidth]{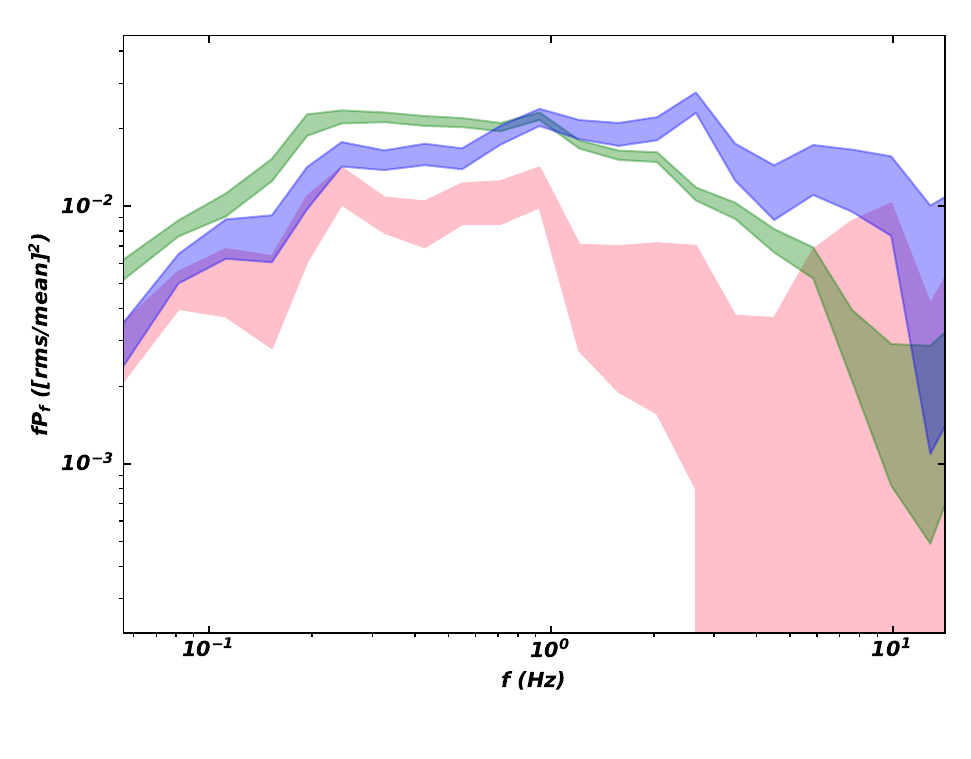}
	\caption{High, Intermediate \& Low band PSDs of the data. The shaded regions are the 1$\sigma$ error regions of the Low (pink), Intermediate (green) and High (blue) energy bands from the data.}
	\label{fig:DATAPSD}
\end{figure}

\section{Modeling}
\label{sec:MODELLING}
In many respects, the model we apply here is the same as that of MD18b; that is, we fit first to the energy spectrum, applying the constraints from this spectral fit to our energy-dependent timing model. We then jointly fit the predictions of this timing model to the power spectra and frequency-dependent time lags of our data. In this section we will outline any differences in prescription for the radial dependence of generated variability, emissivity, viscous timescale, and fluctuation damping from that of the MD18b model. We will also outline the new implementation of the impulse response function, important for the incorporation of disc reflection and thermal reprocessing.

\subsection{Spectral Stratification}
\label{sec:SPECTRALSTRATIFICATION}
In Section~\ref{sec:INTRODUCTION}, we discussed two observational features of hard state BHXRBs that will influence our choice of spectral decomposition. First and foremost are the high-frequency soft lags, potentially due to disc reflection/reprocessing. To test this idea, we must include at least a thermal disc component, and some form of Comptonisation with associated reflection. Second, the often-seen lag between Compton-dominated bands requires that the hot flow be stratified in spectral shape in either time (spectral pivoting) or radius, or both. Previous analysis of Cyg X-1 showed the need for multiple Comptonisation components when considering the Compton lag and the frequency-resolved spectra (MD18a/b, \citealt{AD18}). Here we test this hypothesis for the case of GX 339-4, including reverberation to yield more stringent constraints on the geometry of the inner accretion flow.

\begin{figure}
	\includegraphics[width=\columnwidth]{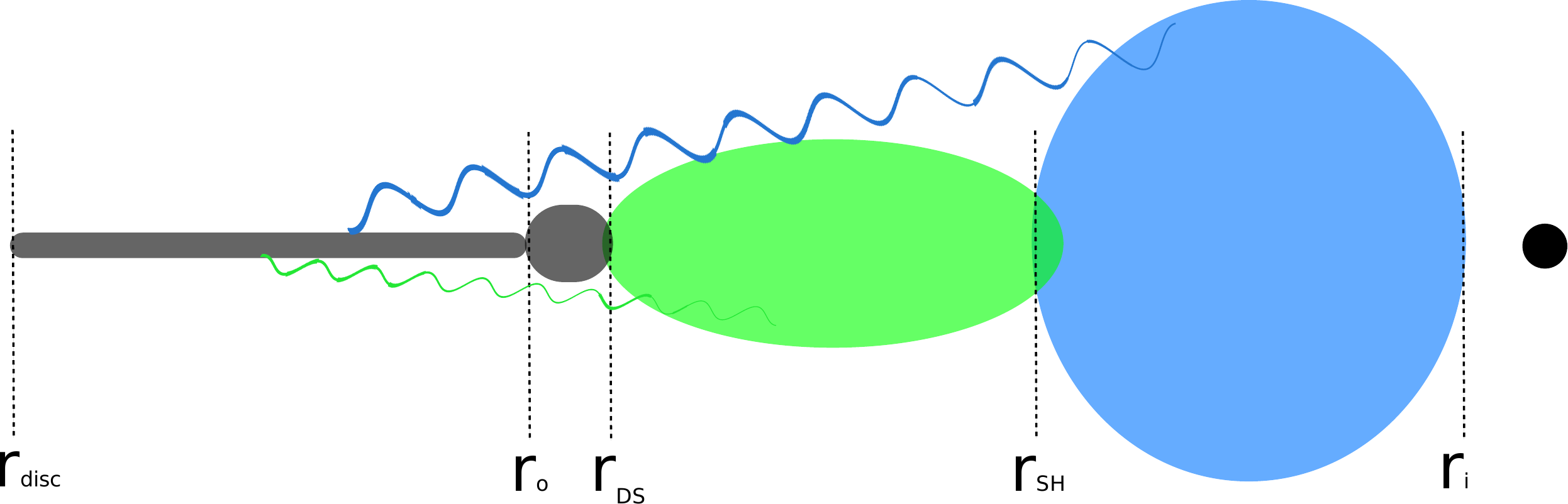}
	\caption{The physical geometry of the flow assumed in our model. The thin grey region between $r_{disc}$ and $r_o$ denotes the thin disc which does not vary intrinsically on fast timescales but does reflect and reprocess Comptonized emission from the inner regions. The thicker, grey region between $r_o$ and $r_{DS}$ denotes the turbulent inner disc which does vary intrinsically on fast timescales due to its interacting with the Comptonising flow. The green region denotes the fast-varying, spectrally \textit{soft} zone, while the cyan region denotes the fast-varying, spectrally \textit{hard} zone. Thermal clump dissipation, disruption by the MRI, or a combination of these effects, would result in damping of fluctuations as they propagate from the \textit{soft} to the \textit{hard} region as required by the data, although our model does not distinguish between these (or other) mechanisms of damping.}
	\label{fig:GEOMETRY}
\end{figure}

We model the time-averaged spectrum with three basic components: a disc blackbody component, $D(E)$, a \textit{soft} Compton component, $S(E)$, and a \textit{hard} Compton component, $H(E)$. We also include the reflection of the \textit{soft} and \textit{hard} Compton emission from the thermal disc, $R_S(E)$ and $R_H(E)$ respectively. Finally we account for thermal reprocessing on the disc, where some component of the blackbody emission, $D(E)$, is made up of thermally reprocessed photons from the Comptonisation. Both the reflected and reprocessed contributions are delayed with respect to their direct Compton components via the Impulse Response Function of Section~\ref{sec:Transferfn}.

The associated geometry is displayed in Fig.~\ref{fig:GEOMETRY}. Here the grey region from $r_{disc}$ to $r_o$ denotes the thermal blackbody-emitting `stable' thin disc, for which we do not model propagating fluctuations since it is assumed that this disc is produces constant emission, as observed in the disc-dominated high/soft state (e.g. \citealt{RM06}). While not intrinsically variable, this material responds to the hard X-ray illumination, producing variable reflected and reprocessed emission. The thicker grey region from $r_o$ to $r_{DS}$ denotes the `variable' disc where turbulence is generated by interaction with the hot flow. This `variable' disc region also contributes to the thermal blackbody component. The green region from $r_{DS}$ to $r_{SH}$ denotes the outer part of the hot flow itself which produces the \textit{soft} Comptonisation components. Finally the blue region from $r_{SH}$ to $r_i$ denotes the inner part of the hot flow which produces the \textit{hard} Comptonisation component. We model the generation and propagation of fluctuations throughout the variable disc and hot flow, from $r_o$ down to $r_i$.

The amount of thermal disc emission resulting from reprocessing of each component is calculated following the \cite{GD14} procedure, where we take the reprocessed luminosity from the \textit{soft} (\textit{hard}) Comptonisation to be
\begin{equation}
\label{eq:LREP}
L_{rep,\,soft\,(hard)} = (\Omega/2\pi)L_{soft\,(hard)} - L_{ref,\,soft\,(hard)},
\end{equation}
where $(\Omega/2\pi)$ is the solid angle subtended by the disc with respect to the hot flow, fixed to that derived from the spectral fit. The fraction of the total thermal disc emission resulting from reprocessing is then $f_{rep} = (L_{rep,\,soft} + L_{rep,\,hard}) / L_{disc}$. Of the remaining disc emission, some will be from the ``stable" disc which produce a negligible variability signature, and some will be from the variable inner disc where interaction with the hot flow is active. We therefore prescribe some fraction of the thermal emission to be variable, denoting this fraction $f_{disc,\, var}$. $f_{disc,\, var}$ is allowed to be a parameter of the timing fit, constrained such that $f_{disc,\,var} < 1 - f_{rep}$. Any remaining contribution to the thermal component must therefore be a constant component, calculated as $f_{disc,\,const} = 1-f_{disc,\,var}-f_{rep}$. By separating our thermal emission into reprocessed and intrinsic fractions, the model is able to treat both propagation of fluctuations from the disc and reverberation on the disc simultaneously. In this simplified picture, reprocessing takes place beyond the disc-flow interaction at $r > r_o$, while the intrinsic, variable emission takes place in the disc-flow interaction zone where the disc begins to be disrupted at $r_o > r > r_{DS}$ (see Fig.~\ref{fig:GEOMETRY}). In reality some thermal reprocessing of hard X-rays will take place in the disc-flow interaction zone, but if this interaction region is relatively small it will subtend only a small solid angle with respect to the illuminating flow, and so the light-travel lags will be dominated by reverberation from further out on the disc. Since the interacting disc zone between $r_o > r > r_{DS}$ is no larger than a few $R_g$ in most of the explored parameter space, this separation of intrinsic and reprocessed disc components is reasonable. We also tacitly assume that all four thermal components - the stable disc, the variable disc and thermalised emission from reprocessing from the \textit{hard} and \textit{soft} Compton spectra - have the same spectral shape. This is unlikely to be strictly true, but the data cannot separately constrain multiple thermal components. We discuss this further in Section 10.

In the model we stratify the interacting disc and hot flow region such that the spectral shape of the emission from each region is one of our three basic components,
\begin{equation}
\label{eq:SPECTRAL}
\bar{F}(E, r_n)=
\begin{cases}
D(E) & \text{if}\ r_n > r_{DS}, \\
S(E) & \text{if}\ r_{SH} < r_n < r_{DS}, \\
H(E) & \text{if}\ r_n < r_{SH},
\end{cases}
\end{equation}
where $r_{DS}$ and $r_{SH}$ are the transition radii between the disc and \textit{soft} Compton, and the \textit{soft} and \textit{hard} Compton regions respectively, indicated in Fig.~\ref{fig:GEOMETRY}. These are analytically derived from the radial scale, the observed spectral components $D(E)$, $S(E)$ and $H(E)$, and the prescribed emissivity (parameterised in Section~\ref{sec:CorrelatedTurbulenceandEmissivity}) such that the luminosity ratios between our three spectral components match those of the integrated emissivity in each region:
\vspace*{5 pt}
\begin{equation}
\label{eq:TRANSITION}
\begin{aligned}
\frac {\int_E f_{disc,\,var}\,D(E) dE}{\int_E S(E) dE} &= \frac {\int_{r_{DS}}^{r_o} \epsilon(r)
	2\pi r dr }{\int_{r_{SH}}^{r_{DS}} \epsilon(r) 2\pi r dr }, \\
\frac {\int_E S(E) dE}{\int_E H(E) dE} &= \frac {\int_{r_{SH}}^{r_{DS}} \epsilon(r)
	2\pi r dr }{\int_{r_i}^{r_{SH}} \epsilon(r) 2\pi r dr }.
\end{aligned}
\end{equation}

\subsection{Correlated Turbulence and Emissivity }
\label{sec:CorrelatedTurbulenceandEmissivity}

There have been a number of complementary proposals for the source of the peaks seen in the power spectra of hard state black hole binaries. Veledina (2016, 2018) provides evidence that the power spectral peaks can be generated by interference of a single variability component which propagates through two distinct emission regions which are separated radially and hence related with a time delay. The mass accretion rate fluctuation first affects Comptonisation in the outer flow due to its effect in changing the disc seed photons, and some time later the same fluctuation enhances Comptonisation of cyclo-synchrotron photons in the inner flow. This gives rise to interference, which they propose to be the origin for the double peaked power spectra. On the other hand, \cite{ID12a} show that enhancement of variability at specific radii can also result in multi-peaked power spectra. \cite{R17a} also proposed that distinct variability timescales in different regions can produce distinct power spectral peaks. These three proposals are not mutually exclusive, and may indeed all be true in some sense. In MD18a/b we therefore developed a formalism which could incorporate all of these effects, by including distinct variability timescales in separate spectral regions, allowing a variability prescription which could enhance the generated variability at particular radii, but also coupling these radii to regions of enhanced emission so that the interference picture of Veledina could be encompassed within the parameter space.

The variability generated at each radius in our model is described by the value $F_{var}(r)$, the fractional variability per radial decade ($F_{var}(r) = \sigma_{rms}(r_n) \sqrt{N_{dec}}$ where $\sigma_{rms}(r_n)$ is the generated root-mean-square variability and $N_{dec}$ is the number of simulated annuli per radial decade, following MD18b). In an attempt to replicate the highly structured, well constrained data of the Cygnus X-1 bright hard state modeled in MD18a/b, a radial variability profile composed of three Gaussian functions of free width, free amplitude and (in all but one case), free position was prescribed. This required an undesirable 8 free parameters, in addition to those required by other aspects of the model. However given that the data we model here features both more noise and less evidence of distinct peaks in the power spectra (Fig.~\ref{fig:DATAPSD}), we simplify this radial variability profile into a sum of a single Gaussian and a radius-independent constant; this profile is analogous to the radially constant turbulence by the magneto-rotational instability (MRI; \citealt{BH98}), with enhancement at some distinct radius:
\begin{equation}
\label{eq:newFvar}
F_{var}(r) = F_{var,\,C} + A_{en} e^{-\frac{r-r_{en}}{2\sigma_{en}^2}}
\end{equation}
with $F_{var,\,C}$, $A_{en}$, $r_{en}$ and $\sigma_{en}$ being model parameters. These parameters are dynamically constrained in the fitting procedure such that for all radii we have $0 < F_{var}(r)/\sqrt{N_{dec}} < 0.33$, the lower bound being obvious as `negative' variability power does not make physical sense, and the upper bound such that the variability generated is not so large that we have negative mass accretion rate at any radius, following tests carried out in MD18b. We also limit $r_{en}$ to lie within the radial range from $r_i$ to $r_o$ to avoid venturing into degenerate parameter space.

As in MD18b, this variability profile is coupled to the emission profile of our interacting disc/hot flow, enhancing the emission in the same radial range as the turbulence is enhanced. However we also include an additive term which depends on radius as a power law with index $\gamma$, a model parameter. This is akin to gravitational dissipation in the thin disc case for $\gamma = 3$, although in the case of the hot flow this index has not been predicted from fundamental accretion theory. The emissivity therefore has functional form
\begin{equation}
\label{eq:newemiss}
\epsilon(r) \propto r^{-\gamma} + Z_{en} e^{-\frac{r-r_{en}}{2\sigma_{en}^2}},
\end{equation}
with additional free parameters $\gamma$ and $Z_{en}$. We note that the radial power law term here is \textit{additive} rather than \textit{multiplicative} as was the case in MD18b. Since we now feature only one Gaussian term in our coupled fractional variability profile, a multiplicative term would be overly restrictive for the emissivity.

\subsection{Propagation Speed}
\label{sec:PropagationSpeed}
Previous propagating fluctuation models (\citealt{AU06}, ID11, MD18a) have prescribed a continuous power-law radial dependence for the viscous frequency, such that $f_{visc}(r) = B r^{-m} f_{kep}(r)$ at all points in the modeled region with $f_{kep}(r)$ being the Keplerian frequency. In MD18b we allowed a more complex viscous profile by assuming distinct power law dependencies in each of the two Comptonisation regions, as it was not clear whether regions of distinct spectra would have the same radial dependence in viscosity. In this case, since we are also modeling the interacting disc region which almost certainly will have a distinct viscous timescale from the hot flow, we assume that the entire Comptonising flow has the same (unknown) radial dependence in viscous frequency, but that this is distinct from the viscous timescale in the variable disc region,
\begin{equation}
\label{eq:2visc}
f_{visc} = 
\begin{cases}
B_{disc} r^{-m_{disc}} f_{kep}(r) & \text{if } r\geq r_{DS} \\
B_{flow} r^{-m_{flow}} f_{kep}(r) & \text{if } r<r_{DS}.
\end{cases}
\end{equation}

The adherence of so many hard state black hole and neutron star sources to the low-frequency-break-QPO relation leads us to assume that the absence of a QPO in this observation is an observational rather than an intrinsic effect, and that the viscosity in our inner disc will still adhere to this relation (ID11). We therefore fix the viscosity in our variable disc to agree with this association, with $B_{disc} = 0.03$ and $m_{disc} = 0.5$ in equation~(\ref{eq:2visc}). In the Comptonising flow on the other hand, we allow the viscosity to have a different radial dependence and amplitude, so that $B_{flow}$ and $m_{flow}$ are model parameters.

\subsection{Damping}
\label{sec:Damping}

As our mass accretion rate fluctuations propagate, we allow for the possibility that they are damped, due to disruption by the MRI, or due to evaporation of accreting clumps of the thermal disc as they propagate through the optically thin flow. As in MD18b, we incorporate this into the model by prescribing damping at the spectral transition radii via
\begin{equation}
\label{eq:DAMPING}
D_{ln} =
\begin{cases}
D_{DS} & \text{if } r_l \geq r_{DS} > r_n > r_{SH}, \\
D_{SH} & \text{if } r_{DS} > r_l > r_{SH} \geq r_n, \\
D_{DS}D_{SH} & \text{if } r_l \geq r_{DS}, r_{SH} \geq r_n, \\
1 & \text{otherwise},
\end{cases}
\end{equation}
for arbitrary annuli $r_l$ and $r_n$.

In MD18b (following \citealt{R17a}), a Green's function term was included describing the impulse response of the flow to the mass accretion rate fluctuations. This term described the predicted smoothing out of mass accretion rate fluctuations due to viscous torques (\citealt{FKR}). However the results of MD18b and preliminary tests on these data indicated that this effect is preferred to be negligible in these models. Those results suggest that either the coherence of mass accretion rate fluctuations is independent of their length scale, or more likely that the propagation lengths in the observed accretion flows are too short for this viscous smoothing to become important. In order to minimize the number of free parameters in our fits, we omit the smoothing effect here.

\subsection{The Impulse Response (Transfer) Function}
\label{sec:Transferfn}

Emission which is reflected from, or reprocessed by, the disc will experience a time delay due the light travel time from the hot corona to the disc. In reality, however, this light illuminates a large radial and angular range of the disc, and the amount of reflection/reprocessing we observe from each part of the disc is dependent on the disc truncation, our inclination to the source, and the scale height of the illuminating flow. This `distributed' delay of the driving (Comptonisation) signal not only delays the reflected/reprocessed emission with respect to the Comptonisation but also smoothes out fluctuations on the fastest timescales. The effect of this illumination distribution of the disc on the timing properties of the reverberated signal is encoded in the impulse response function, $IRF(t)$, or its Fourier transform known as the transfer function, $TF(f)$ (see \citealt{U14} and references therein).

In our model we assume that reflection/reprocessing occurs from the outer edge of the disrupted part of the disc (equivalent to the inner edge of the stable disc) at $r_o$, out to $r_{disc} = 400$. The choice of $r_o$ as the inner edge of the reverberation region was for model practicality, detailed in Section~\ref{sec:SPECTRALSTRATIFICATION}, while the maximum solid angle subtended by the flow beyond $400~R_g$ is selected for uniformity with our spectral reflection models where the outer edge of the disc is also fixed at $400~R_g$. Beyond this radius, the solid angle subtended by our illuminating flow is also small and can be neglected. To construct the transfer function, we adapt the method of \citealt{WH91} which describes the time delay for light reflected from a point on the disc at radius, $r$, from a central source:
\begin{equation}
\label{eq:IRF_element}
\tau = \frac{r}{c}[1-sin\,i\,cos\,\phi],
\end{equation}
where $i$ is the inclination of the axis of the disc to the line of sight and $\phi$ is the azimuthal angle between a point on the disc and the projection of the line of sight onto the disc. Of course the radial/vertical structure of the hot flow itself means that the light travel delay will also be a function of the emission point within the flow. However, given that the distance between the \textit{soft} and \textit{hard} regions of our modeled flow is typically $\lesssim 10~R_g \approx 10^{-4}$~light-seconds, and the fact that the radial range of our disc is large with respect to the hot flow size scale, our assumption of a central illuminating source is reasonable (see e.g. \citealt{GD14}).

Here we have assumed that the IRF is energy independent, on the basis of our assumption that all the thermal components have the same spectral shape. However Comptonized photons incident on a given radius will be reprocessed down to the blackbody temperature at that specific annulus, rather than the peak disc temperature, giving the IRF an energy dependence (see e.g. \citealt{U14}; \citealt{GD17}). While this assumption is necessitated for model practicality here, will assess the effect of this effect in later work with a time-domain simulation of reverberation in our derived geometry. 

\section{Spectral Fit}
\label{sec:SPECTRALFIT}

\begin{figure}
	\includegraphics[width=\columnwidth]{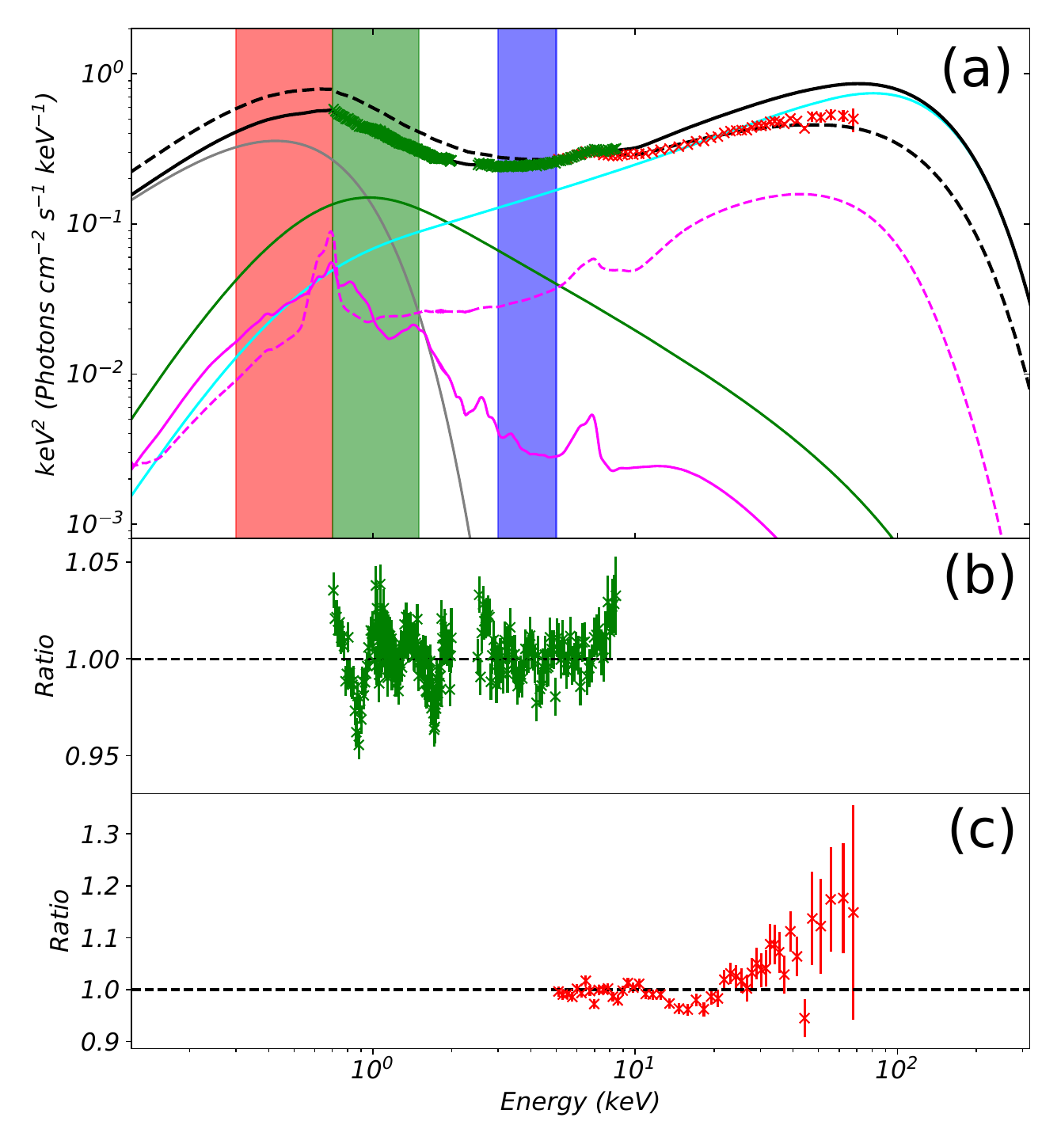}
	\caption{Panel (a): decomposition of O1 into a disc blackbody, two Comptonisation components, and the reflection of these components from the truncated disc. Lines show the total energy spectrum from XMM-Newton (black solid), and the total energy spectrum from NuSTAR (black dashed). For brevity, we display only the components inferred from the XMM fit, but the NuSTAR components feature exactly the same parameters only with the Compton spectral indices swung by $+0.18$ (see Table~\ref{tab:SPECTRAL}). The components shown are: the disc thermal component ($D(E)$, grey solid); the \textit{hard} Compton component ($H(E)$, cyan solid); the \textit{soft} Compton component ($S(E)$, green solid); the disc reflection of the \textit{hard} component ($R_H(E)$, magenta dashed), and the disc reflection of the \textit{soft} component ($R_S(E)$, magenta solid). Crosses show the XMM EPIC-pn (green) and NuSTAR FPMA (red) data. The red, green and blue shaded regions denote the Low ($0.3-0.7$~keV), Intermediate ($0.7-1.5$~keV) and High ($3-5$~keV) energy bands respectively. Panel (b): ratio of XMM-Newton data to model fit. Panel (c): ratio of NuSTAR data to model fit.}
	\label{fig:SED}
\end{figure}

To produce the spectral components which input into our timing fit procedure, we jointly fit simultaneous XMM-Newton spectra ($0.5-10$~keV, ignoring the $2-2.4$~keV region where there are residuals in the response) and NuSTAR ($5-70$~keV) spectra with the model {\tt{tbnew * (diskbb + nthcomp + nthcomp + kdblur * xilconv * (nthcomp + nthcomp))}} in {\sc{xspec}} (version 12.9.1, \citealt{ABH96}; {\tt{tbnew}}, \citealt{WAM00}; {\tt{nthcomp}}, \citealt{ZJM96}). We tie the seed photon temperatures of both Compton components to the inner disk temperature for simplicity, noting however that this would be an inaccurate assumption in the case of cyclo-synchrotron excitation in the inner flow (\citealt{PVZ18}). We also assume that the solid angle subtended on the disc by both the \textit{soft} and \textit{hard} components is the same for model simplicity. For the NuSTAR spectra, we set the spectral indices of our \textit{soft} and \textit{hard} Compton components to be the same as that of the modeled XMM components but swung by $+0.18$, as specified in Table~\ref{tab:SPECTRAL}. This swing is required due to the inherent mismatch in calibration between XMM EPIC-pn data with respect to NuSTAR (and RXTE) The systematic offset is fixed to $+0.18$ here to simplify fitting, although this is slightly more conservative than other XMM-NuSTAR joint studies (e.g. \citealt{I17}). The inclusion of this NuSTAR spectral data is important to constrain the high-energy Comptonisation cutoff in the hard component. This is required since it constrains the total Comptonized luminosity of the source, and therefore the amount of reprocessed emission we should expect from the disc (and conversely how much must be intrinsic to the disc) via equation~(\ref{eq:LREP}).

Due to the mismatch between the NuSTAR and XMM calibrations, it is important to note that at different points in our timing model procedure, we use constraints inferred from either the XMM or NuSTAR spectra. In particular, when calculating energetics - i.e. values inferred from integrals of the spectral components over \textit{all} energies - we use the components inferred from NuSTAR. In equations~(\ref{eq:LREP}),~(\ref{eq:SPECTRAL})~\&~(\ref{eq:TRANSITION}), we therefore use the spectral components inferred from NuSTAR. When computing the \textit{relative} contributions from different components within our three energy bands however (i.e. in the weighting procedure detailed in equations~\ref{eq:weights}~\&~\ref{eq:weights_rev}), we use the XMM-inferred components.

In Fig.~\ref{fig:SED} we show the broadband (de-absorbed) spectral fit to these data. The \textit{soft} (green) and \textit{hard} (cyan) Compton components are produced from the outer and inner regions of the flow respectively, while the reflection from the \textit{soft} component (magenta solid), and the \textit{hard} component (magenta dashed) are also shown. The thermal disc component is shown in grey. As described in Section~\ref{sec:SPECTRALSTRATIFICATION}, some fraction, $f_{rep}$, of this thermal disc component is due to reprocessed emission from the Compton components, which will be delayed with respect to the direct Compton emission in the same way as the reflected emission (by convolution with the impulse response function). $f_{rep}$ is fixed in the timing fits via equation~(\ref{eq:LREP}). There is also thermal disc emission composed of fractions $f_{disc,\,var}$ intrinsic, variable disc emission produced by propagating fluctuations, and $f_{disc,\,const}$ intrinsic, constant disc emission.

\begin{table}
	\begin{tabular}{ccc}
		\hline
		 				& XMM 					&NUSTAR\\
		\hline
		\vspace{+3pt}
		$N_H$ 				& $0.509^{+0.015}_{-0.013}$ 		& ==\\
		\vspace{+3pt}
		$A_{Ne} $ 			& $1.194 ^{+0.056}_{-0.059}$		& ==\\
		\vspace{+3pt}
		$A_{Mg}$ 			& $2.77^{+0.22}_{-0.24}$		& ==\\
		\vspace{+3pt}
		$kT_{in}$ 			& $0.18$ $(F)$				& ==\\
		\vspace{+3pt}
		$n_D$ 				& $5.07^{+0.65}_{-0.62} \times 10^{4}$	& ==\\
		\vspace{+3pt}
		$\Gamma_S$ 			& $2.966^{+0.066}_{-0.062}$		& $3.15^{+0.06{\dagger}}_{-0.07}$\\
		\vspace{+3pt}
		$kT_{e, S}$ 			& $100$ $(F)$				& ==\\
		\vspace{+3pt}
		$n_S$ 				& $0.150 ^{+0.005}_{-0.004}$		&==\\
		\vspace{+3pt}
		$\Gamma_H$ 			& $1.505 ^{+0.097}_{-0.095} $		& $1.685 ^{+0.097{\dagger}}_{-0.095}$\\
		\vspace{+3pt}
		$kT_{e, H}$ 			& $35$ $(F)$				& ==\\
		\vspace{+3pt}
		$n_H$ 				& $0.068 \pm 0.003$			& ==\\
		\vspace{+3pt}
	$\left( \frac{\Omega}{2\pi}\right)$ 	& $-0.297 \pm 0.022$			& ==\\
		\vspace{+3pt}
		log($\xi_i$) 			& $3.072^{+0.032}_{-0.029}$		& ==\\
		\hline
		$\chi^2/dof$ 			& \multicolumn{2}{c}{$2902.3/2469$ (combined)} \\	
		\hline
		\multicolumn{3}{l}{\textsuperscript{$\dagger$}\footnotesize{These are tied to the XMM parameters so that }}\\
		\multicolumn{3}{l}{\footnotesize{$\Gamma_{NuSTAR} = \Gamma_{XMM}+0.18$.}}\\
	\end{tabular}
	\caption[caption]{Parameter results of spectral fitting to O1 using the model {\tt{tbnew * (diskbb + nthcomp + nthcomp + kdblur * xilconv * (nthcomp + nthcomp))}}, fit simultaneously to XMM and NuSTAR data.}
	\label{tab:SPECTRAL}
\end{table}

\section{Timing Fit Procedure}
\label{sec:TimingFit}
In all modeling we assume that the central black hole has a mass of $7~M_{\odot}$ and that we are inclined by $50^{o}$ to the system, bearing in mind however that these parameters could in reality be different but correlated (i.e. higher mass and lower inclination or vice-versa; see Fig.~7 of \citealt{HJTC17}). The propagating fluctuations model uses $N_r = 50$ radial bins, reduced from $N_r=70$ in MD18b due to the smoother structure in our data here making higher radial resolution unnecessary.

Following the procedure for extracting power spectra and time lags from our model outlined in Appendix~\ref{Timing Formalism}, we simultaneously fit the power spectra in each energy band, and time lags between each distinct pair of energy bands by minimizing:
\begin{align*}
\chi^2 = \sum^J_{j=1} \left\{ \frac{[P^{\,mod}_{L}(f_j)-P^{\,obs}_{L}(f_j)]^2}{\Delta P^{\,obs}_{L}(f_j)^2} + \frac{[P^{\,mod}_{I}(f_j)-P^{\,obs}_{I}(f_j)]^2}{\Delta P^{\,obs}_{I}(f_j)^2} + \right. \\ \left. \frac{[P^{\,mod}_{H}(f_j)-P^{\,obs}_{H}(f_j)]^2}{\Delta P^{\,obs}_{H}(f_j)^2} + \frac{ [\tau^{\,mod}_{LH}(f_j)-\tau^{\,obs}_{LH}(f_j)]^2}{\Delta \tau^{\,obs}_{LH}(f_j)^2} + \right. \\
\left. \frac{ [\tau^{\,mod}_{IH}(f_j)-\tau^{\,obs}_{IH}(f_j)]^2}{\Delta \tau^{\,obs}_{IH}(f_j)^2} + \frac{ [\tau^{\,mod}_{LI}(f_j)-\tau^{\,obs}_{LI}(f_j)]^2}{\Delta \tau^{\,obs}_{LI}(f_j)^2} \right\},
\end{align*}
where superscripts $mod$ and $obs$ denote the modeled or observed statistics respectively.

\section{Timing Fit}
\label{sec:TIMINGFIT}

\begin{figure}
	\includegraphics[width=\columnwidth]{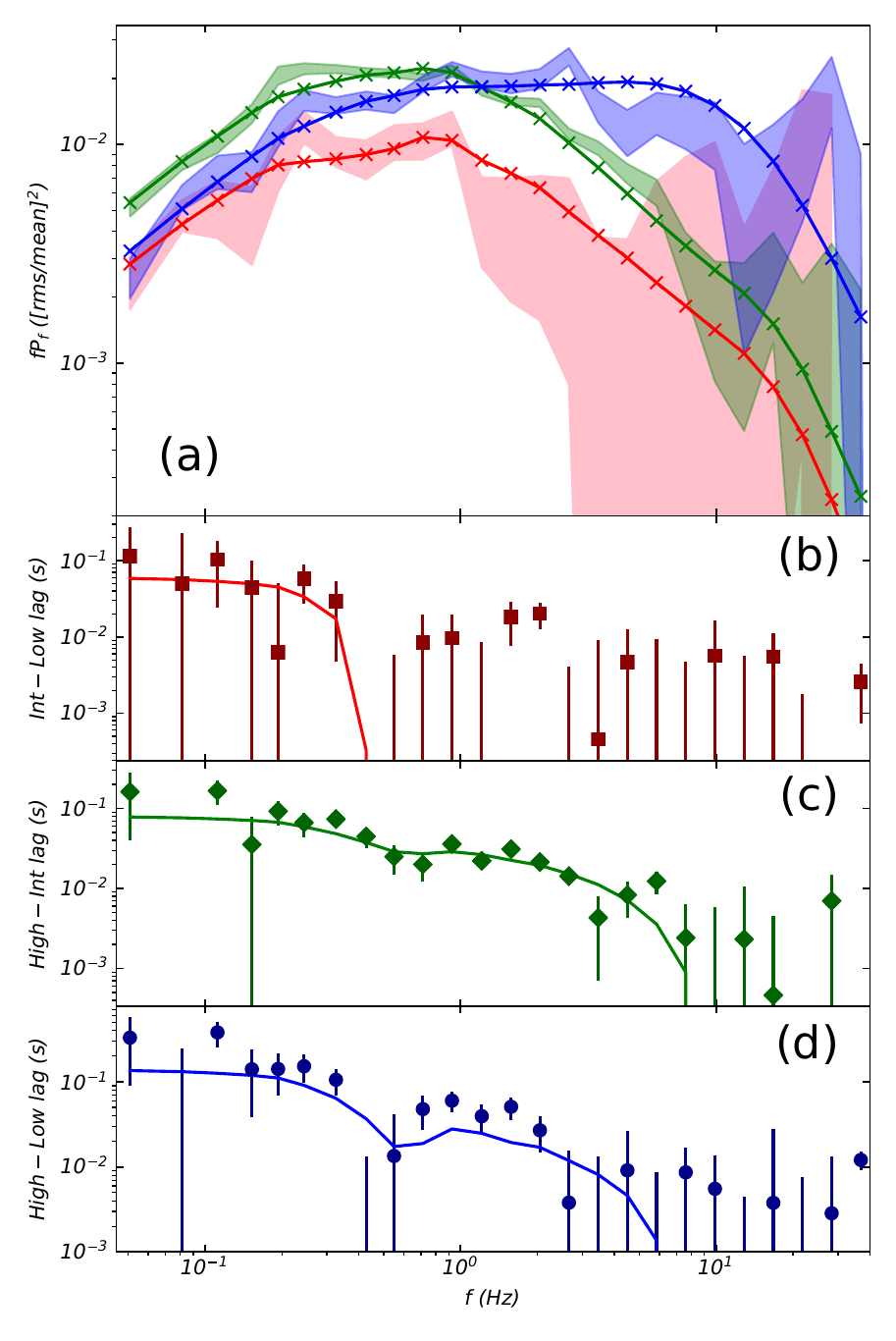}
	\caption{Timing fit using spectral model of Fig.~\ref{fig:SED}. Panel (a): High, Intermediate \& Low band PSDs. The shaded regions are the 1$\sigma$ error regions of the Low (pink), Intermediate (green) and High (blue) energy bands from the data. The solid barbed lines show the Low (red), Intermediate (green) and High (blue) energy model outputs. Panel (b): Time lags between Low and Intermediate energy bands. Panel (c): Time lags between Intermediate and High energy bands. Panel (d): Time lags between Low and High energy bands. In all time lag panels, the symbols denote the data, and solid lines denote the model.}
	\label{fig:PSDLAG}
\end{figure}

\begin{table*}
	\centering
	\begin{tabular}{cccccccc}
		\hline
		\vspace{+3pt}
		$B_{disc}$ & $m_{disc}$ & $B_{flow}$ & $m_{flow}$ & $r_o$ & $r_i$ & $Z_{en}$ &$r_{en}$ \tabularnewline
		$0.03\,(F)$ & $0.5\,(F)$ & $0.175^{+0.039}_{-0.041}$ & $1.20 \pm 0.14$ & $19.5^{+3.2}_{-2.3}$ & $4.02^{+0.29}_{-0.46}$ & $6.01^{+0.67}_{-0.56}\times 10^{-5}$ & $19.2^{+0.4}_{-3.4}$ \tabularnewline
		\hline
		\vspace{+3pt}
		$\sigma_{en}$ & $\gamma$ & $F_{var, C}$ & $A_{en}$ & ${D}_{DS}$ & ${D}_{SH}$ & $f_{disc,\,var}$ & $\boldsymbol{\chi^2/dof}$ \tabularnewline 
		$0.246\pm0.037$ & $4.92^{+0.56}_{-0.11}$ & $1.100^{+0.055}_{-0.017}$ & $1.64^{+0.69}_{-0.21}$ & $1.55^{+0.10}_{-0.43}$ & $4.91^{+0.95}_{-0.04}$ & $0.21\pm 0.05$ & $\boldsymbol{462.9/276}$ \tabularnewline
		\hline
	\end{tabular}
	\caption[caption]{Spectral-timing parameter results of fits to O1 assuming the spectral fit of Fig.~\ref{fig:SED}, with parameter definitions described throughout Section~\ref{sec:MODELLING}. Relevant radii not shown in the table as they were fixed, or derived from the model, include: $r_{disc} = 400$; $r_{DS} = 18.6$; $r_{SH}= 6.6$.}
	\label{tab:TIMING}
\end{table*}

\begin{figure}
	\includegraphics[width=\columnwidth]{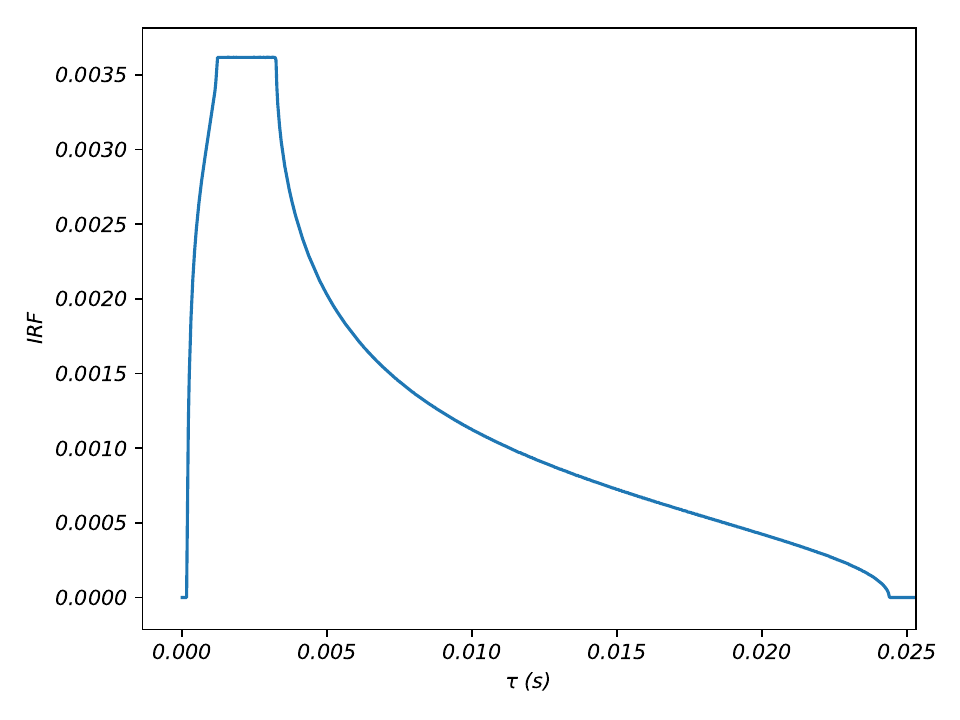}
	\caption{Impulse response function of the disc required for fit of Fig.~\ref{fig:PSDLAG}, inferred from parameter $r_o$ and equation~(\ref{eq:IRF_element}).}
	\label{fig:IRF}
\end{figure}

\begin{figure}
	\includegraphics[width=\columnwidth]{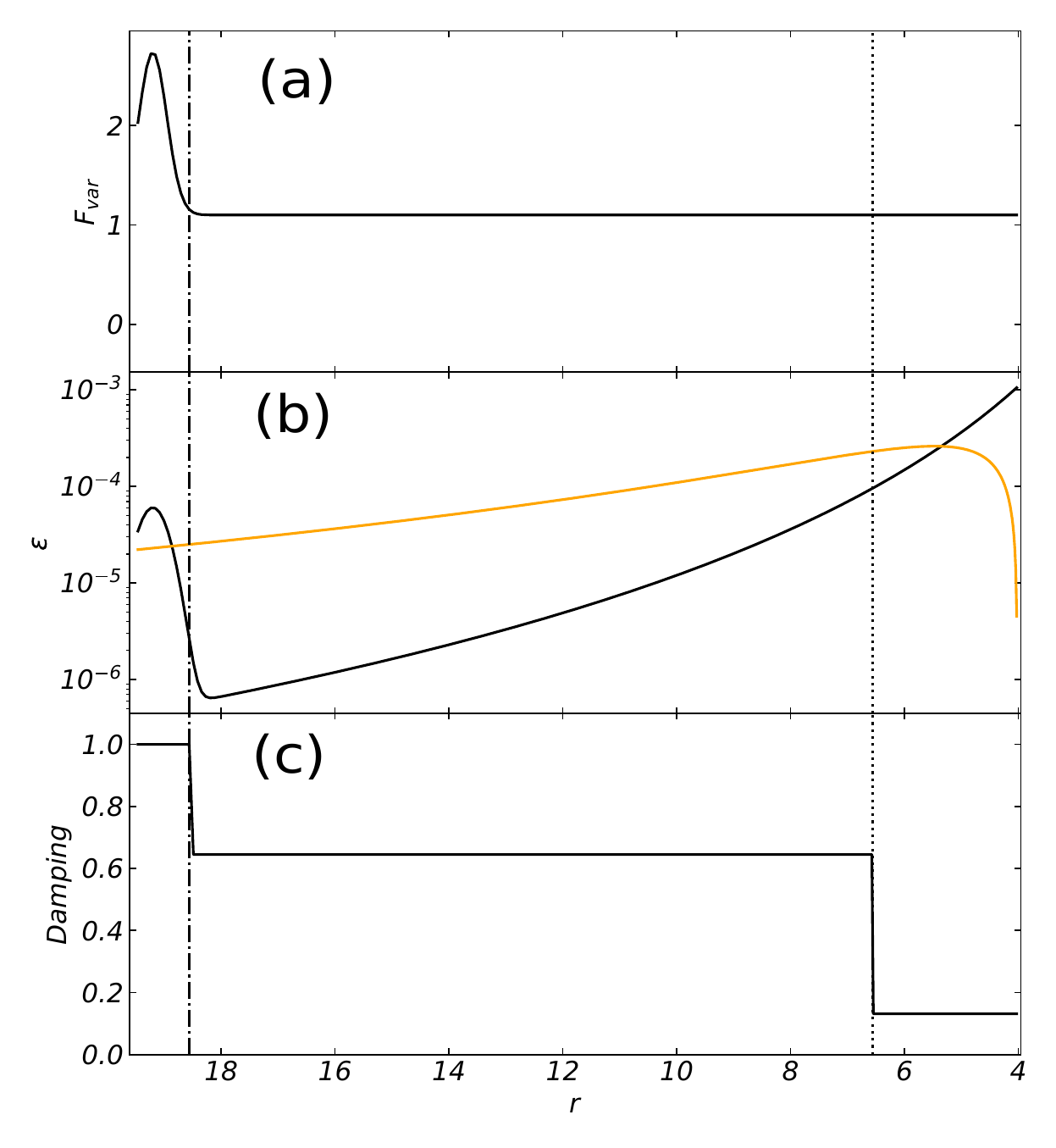}
	\caption{Vertical dot-dash line and vertical dotted lines denote $r_{DS}$ and $r_{SH}$ respectively. Panel (a): Fractional variability ($F_{var}$) profile required for fit of Fig.~\ref{fig:PSDLAG}, inferred from equation~(\ref{eq:newFvar}) and parameters $F_{var,\,C}$, $A_{en}$, $\sigma_{en}$ and $r_{en}$. Panel (b): Black, solid line denotes emissivity ($\epsilon$) profile required for fit of Fig.~\ref{fig:PSDLAG}, inferred from equation~(\ref{eq:newemiss}) and parameters $\gamma$, $Z_{en}$, $\sigma_{en}$ and $r_{en}$. Orange solid line denotes Novikov-Thorne form $\epsilon (r) \propto r^{-3} \left(1-\sqrt{r_i/r}\right)$ profile for comparison. Panel (c): Fluctuation damping profile required for fit of Fig.~\ref{fig:PSDLAG}, inferred from parameters $D_{DS}$ and $D_{SH}$. Propagated fluctuations from outer regions are multiplied by this factor as they pass into interior regions (i.e. fluctuations from $r>r_{DS}$ are multiplied by $1/D_{DS} = 0.65$ as they pass $r_{DS}$, fluctuations from $r>r_{SH}$ are multiplied by $1/D_{SH} = 0.2$ as they pass $r_{SH}$; see equation~(\ref{eq:DAMPING}) and Appendix~\ref{Timing Formalism} for details).}
	\label{fig:PROFILES}
\end{figure}

Fig.~\ref{fig:PSDLAG} shows the optimal joint fit to the PSDs and lag-frequency spectra of our three bands, obtained using the constraints of the above spectral model. The parameters of the timing fit are shown in Table~\ref{tab:TIMING}, along with the combined $\chi^2/dof$. This combined reduced chi-squared is an average of $\{\chi^2/dof\}_{PSDs} = 202.3/138$ and $\{\chi^2/dof\}_{lags} = 260.6/138$; a slightly better fit to the power spectra than to the lags. Unlike the bright hard state data of MD18b, the model reproduces the power spectral features of the data quite well in this instance. Quantitatively, this fit shows a sum of ratio residuals in the power spectra which is a factor 1.5 smaller than that of MD18b over the same number of frequency bins, even with a significantly simpler emission/variability profile in this case (5 fewer free parameters). To check consistency between the spectral and timing fits, the inferred inner disc radius of $19.5$ was plugged back in to the {\tt{kdblur}} component of the spectral model which originally used an inner disc radius of $30$~$R_g$; we find a change in reduced chi squared of only $\Delta \chi^2_{red, spec} = 0.01$, so this would not affect our results noticeably. The inferred IRF is shown in Fig.~\ref{fig:IRF}. The inferred variability and emission profiles are shown in Fig.~\ref{fig:PROFILES}~a~\&~b, where the abscissa extends from the inner edge of the stable disc at $r_o = 19.5$ to the inner edge of the hot flow at $r_i = 4$. Here we see a constant generated fractional variability associated with an $\sim r^{-4.9}$ emissivity dependence in most of the flow, with both profiles also accompanied by an enhancement of turbulence/emission in the $r_o$-$r_{DS}$ region, as we might expect for a highly unstable disc/flow layer. Similar to the bright hard state of Cygnus X-1 (MD18b), significant damping of fluctuations is required between distinct Comptonisation zones in order to reproduce the change in relative power spectral amplitudes at different frequencies (Fig.~\ref{fig:PROFILES}c). The switch in dominance in the power spectra occurs in the correct positions when the hard band dominates over the intermediate band at higher frequencies, while the low-energy variability power is suppressed at all frequencies. The shape of the lag-frequency spectra is less obvious due to their similar amplitudes at many frequencies, but the rough structure between all three bands is approximated by the model.

Our best fit model yields an inner stable disc radius of $r_o \sim 20$, enhanced turbulence in the outer flow/interacting disc, strong damping as we go from the \textit{soft} to \textit{hard} Compton region. These results may be affected by the poorer signal-to-noise compared to the Cygnus X-1 data of MD18a/b which may prevent us from resolving more structure in the power spectra and lag-frequency spectra. However, we can use the inferred best fit parameters to make predictions for the lag-energy spectra and compare these to the data. According to the interpretation given in DM17, the reverberation lag should be evident in the high frequency lag-energy spectra. Making this comparison to the data will help to determine whether we have truly resolved a reverberation lag in this source.

\section{Lag vs. Energy Predictions}
\label{sec:LagE}

\begin{figure}
	\includegraphics[width=\columnwidth]{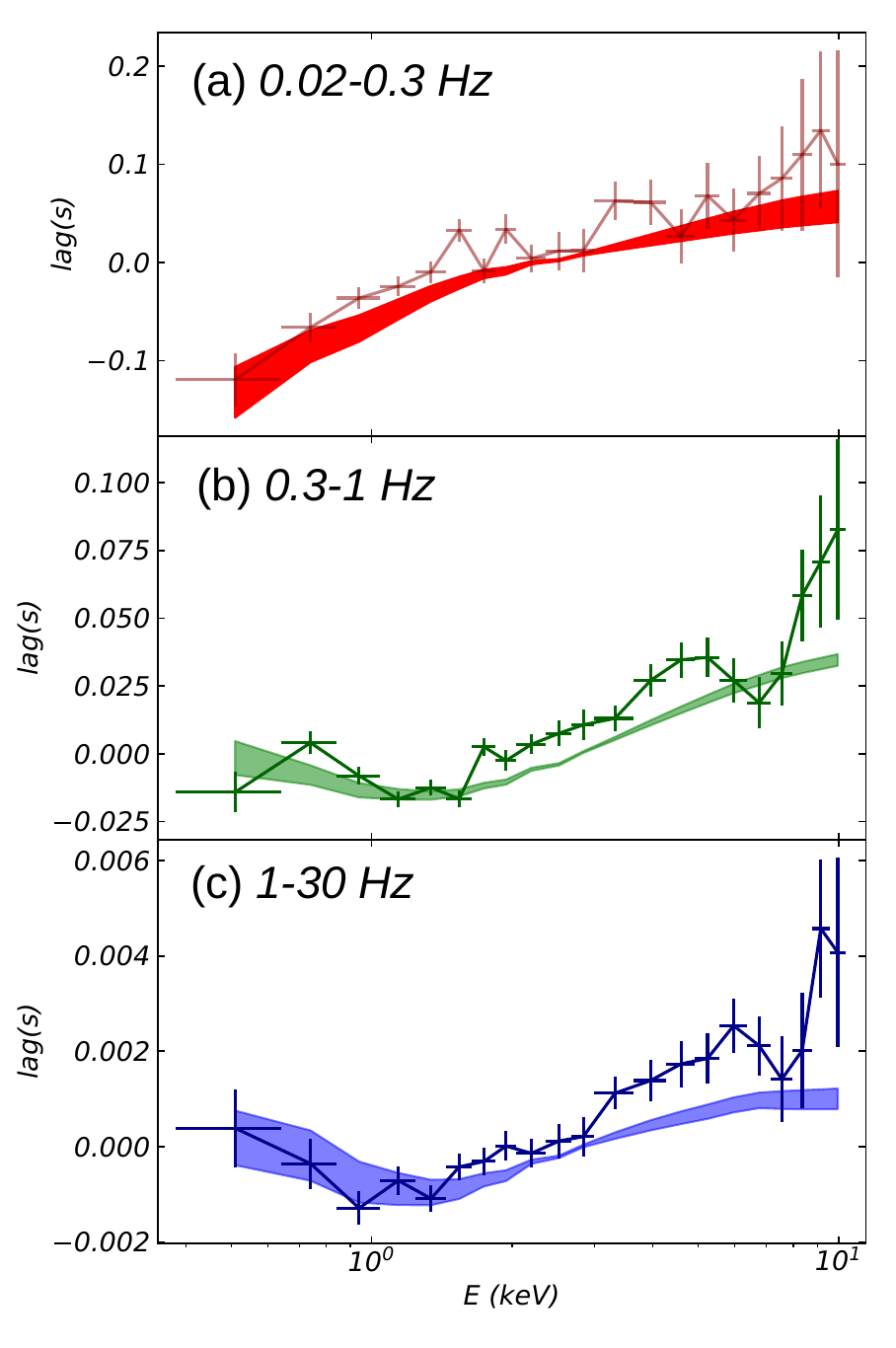}
	\caption{Lag-energy spectra of the data and model in the frequency ranges $0.02-0.3$~Hz (panel (a), red), $0.3-1$~Hz (panel (b), green) and $1-30$~Hz (panel (c), blue). These lags are computed between adjacent small energy bands and a broad reference band ($0.5-10$~keV). The error bars denote the data, and the shaded regions denote the model predictions based on the fit to the PSDs and lag-frequency spectra of Fig.~\ref{fig:PSDLAG}.}
	\label{fig:LAGE}
\end{figure}

We now compare the predictions of the model for the lag-energy spectra, using the parameters established from the fit to the power spectra and lags. We compare these to the data in the three key frequency ranges shown in DM17. We emphasize that these statistics are not a result of direct fits; they are simply predicted by the model. In order to calculate the lag-energy spectra for the data and model, we compute the cross-spectra between the lightcurves in a reference band (from $0.5-10$~keV), and 21 distinct energy bins. From these, lags are computed per the procedure of \cite{U14} in three frequency ranges: $0.02-0.3$~Hz, $0.3-1$~Hz and $1-30$Hz.

From Fig.~\ref{fig:LAGE}a \& b we see that the log-linear trend and magnitude of the $0.02-0.3$~Hz and $0.3-1$~Hz lag-energy spectra observed in the data is well reproduced by the model, correctly describing the range of lags most likely dominated by propagation of mass accretion rate fluctuations through the inner disc and hot flow. Most remarkably however, is the inversion of the $1-30$~Hz lag-energy spectrum at $\sim 1$~keV, indicating the introduction of the soft-lagging component due to thermal reprocessing below $1$~keV. A simple repeat of the calculation of the high-frequency lag-energy spectrum of the model with reprocessing/reflection turned \textit{off} shows no such inversion feature, and a simple log-linear trend down to $-4$~ms lag. The fact that our model includes both propagation and reprocessing within a physically likely spectral decomposition and viable geometry therefore strongly suggests that the observed feature here is indeed thermal reprocessing on the background of a strong propagating-fluctuation signal found within a disrupted disc/hot Comptonising flow structure.

However, the model misses the  statistically significant dip in the lag around the iron line band in the $1-30$~Hz lag-energy spectrum (and to a lesser degree, in the $0.3-1$~Hz lag-energy results also). This dip between $6-8$~keV is notable as it sits close to the FeK$\alpha$ line near $6.7$~keV, suggesting that this complexity is connected to the spectral-timing properties of the reflection spectrum. One key phenomenon that could explain this behaviour is the changing ionization state of the reflector on fast timescales, and we briefly explore this in the following section.

\section{The Varying Ionization State of the Reflector}
\label{sec:Ionisation}

\begin{figure}
	\includegraphics[width=\columnwidth]{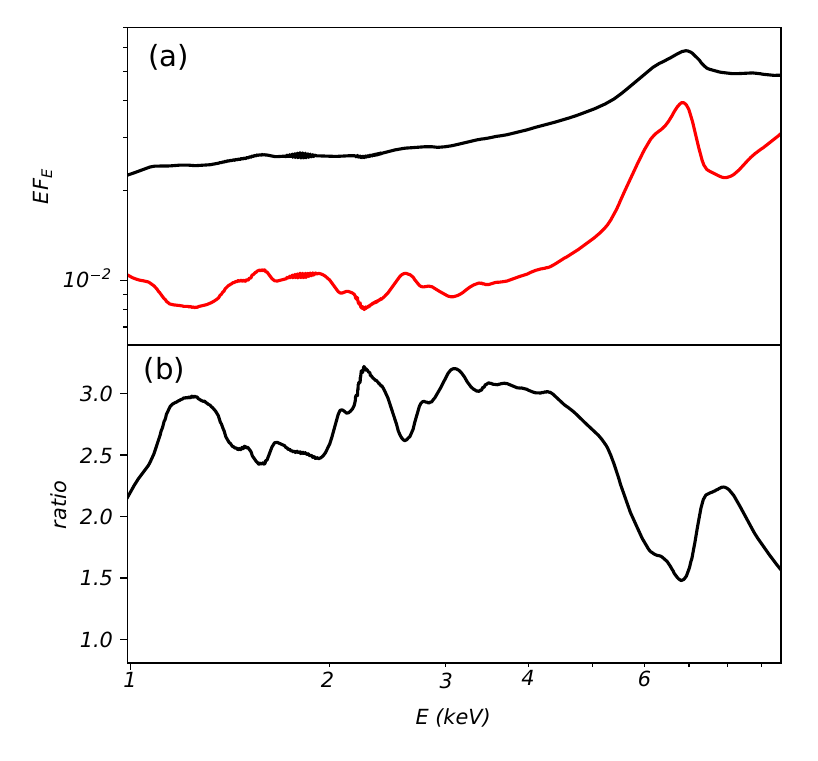}
	\caption{Panel (a): The reflection spectrum before and after shifting the ionization state down by a factor 2, (i.e. the effect on the reflection spectrum if the flux incident on the disc drops by a factor of two, corrected for the resultant reflected flux change itself.) Black line shows the mean ionization state case (i.e. the same as in Fig.~\ref{fig:SED}; the red line shows the same but with ionization state lower by factor 2. Panel (b): The ratio of the spectra in panel (a), demonstrating the change a difference in ionization state can make to the variability properties. Note the dip near the FeK$\alpha$ line.}
	\label{fig:IONIZATION_example}
\end{figure}

\begin{figure}
	\includegraphics[width=\columnwidth]{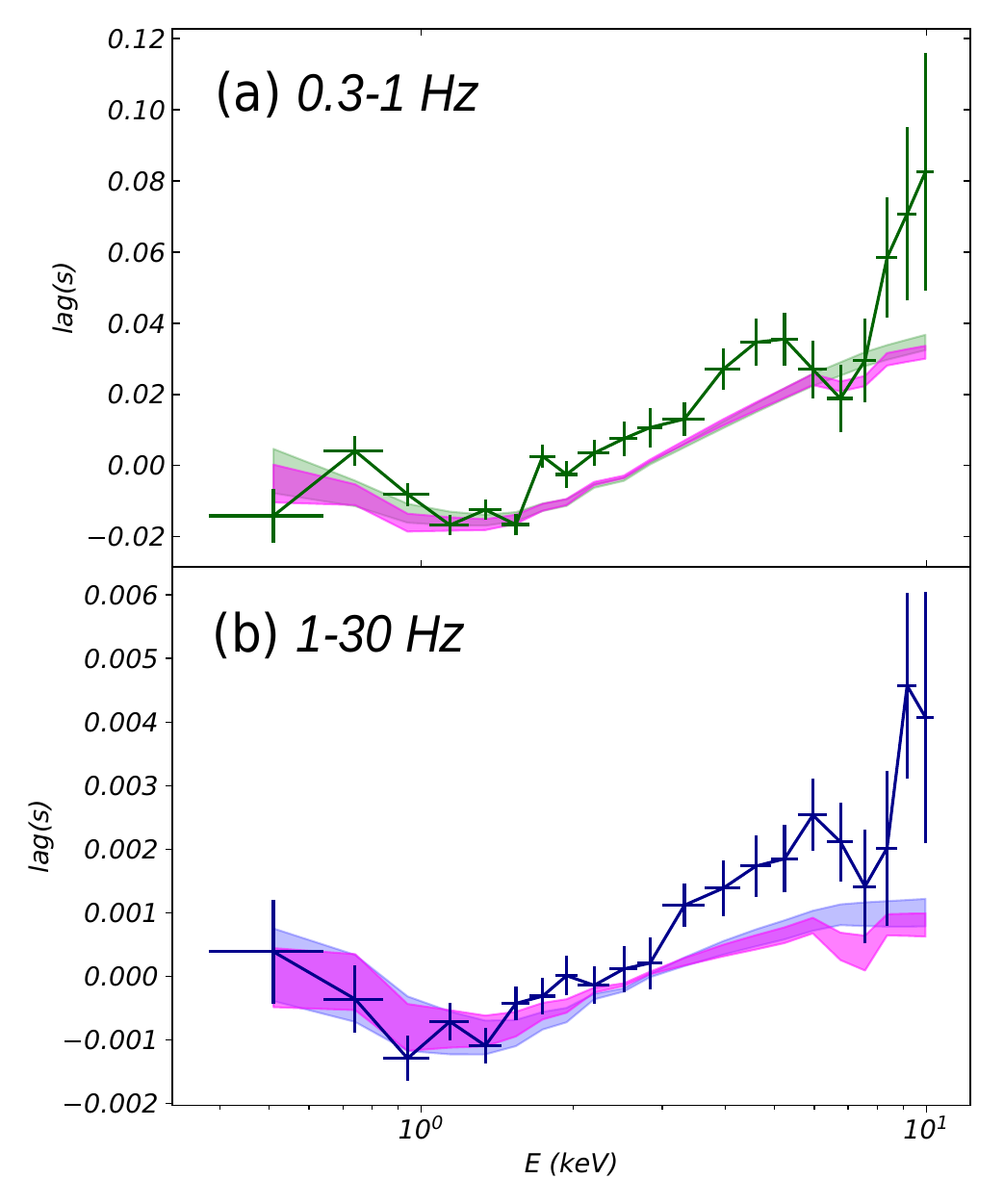}
	\caption{Comparison with modified lag-energy model predictions which has additional modeled variability amplitude above and below the FeK$\alpha$ line owing to variations in the ionization state (see text). Panel (a): Lag-energy prediction for the $0.3-1$~Hz variability, with data show as green error bars, old (constant ionization) model prediction as green, shaded region, and new prediction including variability due to ionization in the magenta, shaded region. Panel (b): Lag-energy prediction for the $1-30$~Hz variability, with data show as blue error bars, old (constant ionization) model prediction as blue, shaded region, and new prediction including variability due to ionization in the magenta, shaded region.}
	\label{fig:IONIZATION}
\end{figure}

An important consideration which our models have so far not included is the effect of the rapidly changing illuminating flux on the ionization state of the disc. Fluctuations in the illuminating flux will produce a correlated variation in the disc ionization state, and the high density of the disc means that the ionization/recombination timescale is very fast so that this tracks the illumination. This ionization state change affects the shape of the reflected spectrum. Fig.~\ref{fig:IONIZATION_example}(a), shows the mean reflected spectrum in our model (black line) compared to the same amount of reflection from a disc of ionization parameter which is a factor 2 lower (red line). The change in flux on the normalization of the reflected emission has been accounted for, this is only 
showing the effect of a change in ion populations. Fig.~\ref{fig:IONIZATION_example}(b) shows the ratio between these two reflected spectra. Plainly there is enhanced variability above and especially below the iron line energy, by a factor $\sim 2.8$ in the $1-5$~keV band, and a factor $\sim 2$ at $8$~keV. For an illuminating flux, $F_{ill}$, we model the fractional change in reflected flux as $(1+f_{ion})\frac{\delta F_{ill}}{F_{ill}}$, while the flux from the iron line itself varies by $\sim \frac{\delta F_{ill}}{F_{ill}}$ only. Fig.~\ref{fig:IONIZATION_example} shows that at these ionization states we can assume $f_{ion}=2$. The effect of this typical $f_{ion}$ on the $0.3-1$~Hz and $1-30$~Hz predicted lag-energy spectra is shown in Fig.~\ref{fig:IONIZATION} (magenta shaded regions in both panels), where the model now exhibits the characteristic dip in lag near $\sim 8$~keV observed in the data. The effect on the other timing statistics is negligible so we do not show them.

We stress that this behaviour is produced in our model because of the enhanced variability of the reflected continuum \textit{around} the Fe-K$\alpha$ line, rather than being produced by a suppression of the line response. This result indicates that if the complex, high-frequency characteristics of the X-ray emission are to be modeled completely, the ionizing effect of the incident emission - and the response timescale of the disc material at different energies - cannot be ignored (see also \citealt{CY12}).

\section{A Higher Frequency Prediction}
\label{sec:PREDICTION}

\begin{figure}
	\includegraphics[width=\columnwidth]{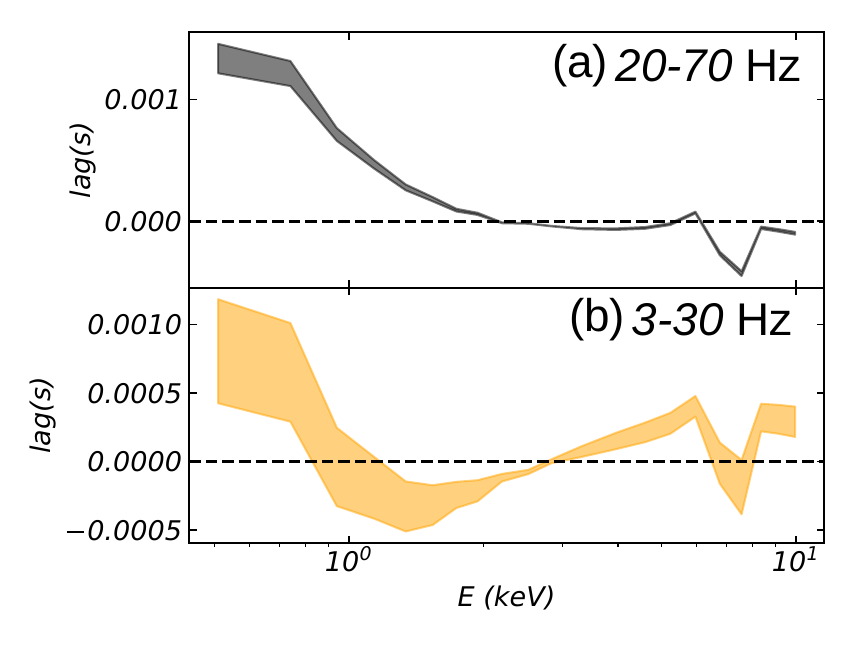}
	\caption{Panel (a): Lag-energy prediction (including additional variability due to ionization changes) in the range $20-70$~Hz. This is a much higher frequency range than that observable by XMM. Panel (b): Lag-energy in the range $3-30$~Hz, to match Fig. 3 of Kara et al. (2019), with which we see qualitative agreement in terms of the lag amplitudes and energetic separation.}
	\label{fig:PREDICTION}
\end{figure}
Our model is now able to approximate the real lag-energy data at frequencies up to $30$~Hz. The XMM-Newton data above this frequency have only limited statistics, but the model can be extrapolated to these higher frequencies to predict the lag-energy spectrum where it is entirely dominated by reverberation. In Fig.~\ref{fig:PREDICTION}(a) we therefore show the predicted lag-energy spectrum of these data for $20-70$~Hz. Clearly, soft-lags now dominate the lag-energy spectrum in exactly the opposite sense to the low-frequency hard propagation lags. We expect just this behaviour for reprocessing of hard X-rays from the fastest hot flow variability, where fluctuation propagation has little influence since these fluctuations are generated closest to the innermost edge of the flow.

While XMM-Newton cannot provide good statistics, we note that the Neutron star Interior Composition ExploreR instrument (NICER; \citealt{NICER12}) has more effective area than XMM-Newton at low energies. Recent work by \cite{K19} presents NICER observations of the $\sim 10$~$M_{\odot}$ XRB MAXI J1820+070 during its fast rise to outburst. Their lowest luminosity spectrum appears to have similar properties to the GX339-4 data shown here, so we show our predicted lag-energy for their 3-30~Hz frequency range in Fig.~\ref{fig:PREDICTION}(b). We note that this has similar structure to their data (see Fig.~3 of \citealt{K19}). This suggests that the truncated disc framework we have described here may be able to successfully model such data up to high frequencies. Their reverberation signature clearly shifts to higher frequencies for higher luminosity/steeper spectra. This is easily explained in the truncated disc model by the disc extending closer to the black hole, as the hot flow shrinks. \cite{K19} also require that the hot flow shrinks, although they extend the Comptonising region vertically rather than radially.

\section{Discussion}
\label{sec:Discussion}
In all hard state sources we see a low-frequency break in the PSD. Many of these also show a low-frequency QPO, where the QPO frequency moves with the low-frequency break. There is now strong evidence that Lense-Thirring precession of the hot flow is the origin of the QPO (\citealt{I16}). This gives us the outer radius of the hot flow, and hence the low-frequency break sets the viscous timescale from this radius. While we do not observe a QPO in these data, we do observe a low-frequency break. We therefore assumed that the material in the interacting disc region adheres to the viscous timescale derived from the $f_{qpo}-f_{lb}$ relation of ID11 ($f_{visc} = 0.03r^{-0.5}f_{kep}(r)$ for $r>r_{DS}$). This relation predicts a truncation radius in this state of $\sim 20$~$R_g$, and this is consistent with the reverberation results predicting $19.5$~$R_g$. This adds to the weight of evidence for Lense-Thirring precession setting the frequency of the QPO.

The astute reader may note that the stable disc truncation radius we have found here, $r_o = 19.5^{+3.2}_{-2.3}$, is a factor $\sim 4$ smaller than that inferred for this observation from the estimate of DM17. However, the calculation in DM17 was a back-of-the-envelope estimate. Our spectral-timing model includes both the underlying propagation lags as well as the light travel paths of the Comptonized photons to the disc, thus providing a more robust estimate of the disc truncation. In the modeling we have performed here, we have spread our reverberation signal across the disc according to an (albeit energy-independent) impulse response function, which results in our lower inferred truncation radius, which is also set in part by the low-frequency break in the power spectrum. Plainly, changing the assumed inclination and mass of GX 339-4 (i.e. moving it within the degenerate parameter space permitted in \citealt{HJTC17}) would also move this estimate. Nonetheless this shows that consideration of the impulse response function for a radially extended disc can result in inferred truncation radii which differ by a factor of a few from commonly-used simpler estimates.

\begin{figure}
	\includegraphics[width=\columnwidth]{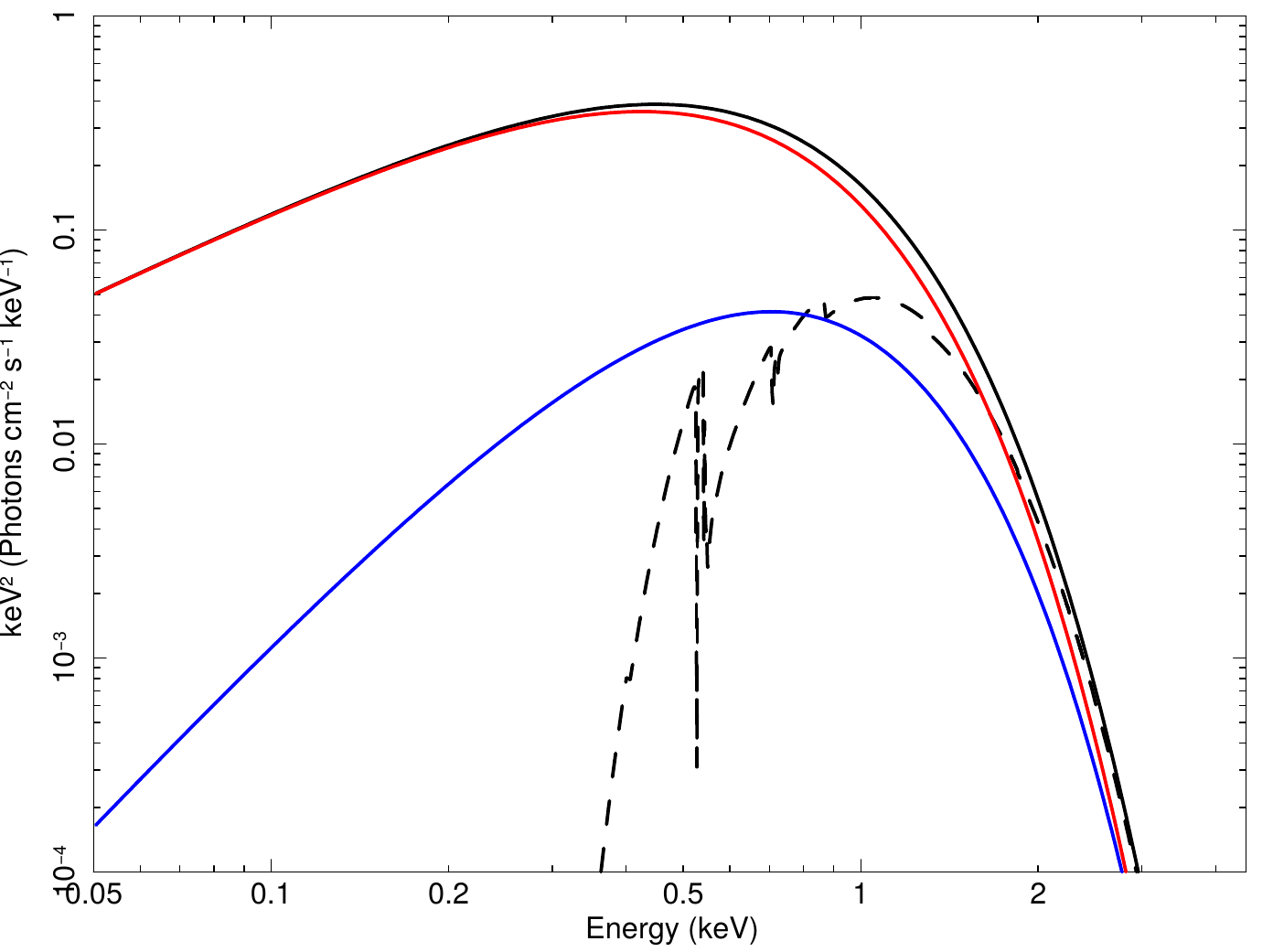}
	\caption{The effect upon the overall thermal spectrum of considering the additional emission produced by the turbulent disc region. The red solid line denotes the {\sc{xspec}} {\tt{diskbb}} spectrum using parameters of Table~\ref{tab:SPECTRAL}. The blue solid line denotes a {\tt{bbodyrad}} spectrum with the same temperature as the {\tt{diskbb}} component, but with $20\%$ of the power, inferred from the fit parameter of $f_{disc,\,var}=0.21$. The black solid line denotes the sum of these two components. The black dashed lined denotes the absorbed total spectrum. We see that, when absorption is accounted for, the difference in shape arising from the {\tt{bbodyrad}} component is inconsequential for the observable energy range.}
	\label{fig:diskvsdiskbbody}
\end{figure}

We have made several assumptions in making the model for the thermal reverberation signal. When performing our spectral fitting, we have assumed that both the turbulent and stable disc regions can be modeled together as a single {\tt{diskbb}} component, with the turbulent disc contributing some mean fraction to the total thermal luminosity in the range measured in the timing analysis. A more physical model would have the turbulent region produce an additional, variable blackbody component on the inner edge of the disc. Fig.~\ref{fig:diskvsdiskbbody} shows a simple xspec model of {\tt{diskbb + bbodyrad}}, where the inner stable disc {\tt{diskbb}} temperature and {\tt{bbodyrad}} temperature are set to be the same ($kT=0.18$~keV as in the model fit). Since $f_{disc,\,var} = 0.21$ in the fit, we set the {\tt{bbodyrad}} component to have 21\% of the total luminosity of the {\tt{diskbb}} component. We see that the blackbody component mildly concentrates the propagating, slow variable emission component towards the highest disc temperatures. However, the black dotted line in Fig.~\ref{fig:diskvsdiskbbody} shows the effect of interstellar absorption on the total spectrum, using {\tt{tbnew}} with the spectral fit parameters. Plainly this severely limits the sensitivity of these data, but we note that lower absorption columns for other objects may make this more visible (e.g. MAXI J1820+070 in \cite{K19}, where the column is only $1.5\times 10^{21}$~cm$^{-2}$).

Placing our resultant truncation radius in the context of other studies, the same NuSTAR dataset is fit by \cite{WJ18} as `Obs 1 2015', though they use the simultaneous Swift XRT data to extend this down to lower energies rather than the higher signal-to-noise (but possibly worse cross-calibrated) XMM-Newton data. They fit the time averaged spectrum with a simpler continuum model, with only a single Comptonisation component (rather than two as used here). Their derived relativistic smearing requires a disc with extreme iron overabundance which extends down to around $2.5~R_g$, an order of magnitude smaller than the radius derived here. If we remove one of our Compton continua, we find a similar fit from our XMM-Newton data, which also requires an extreme iron overabundance of $A_{Fe}=8.67$, and a small innermost disc radius of $2.69~R_g$ (for details see Appendix~\ref{Single_vs_double_Compton}). This fit is considerably worse (combined $\Delta\chi^2>300$) than the two Compton continuum model used in our spectral fits, where the reflection spectrum is solar abundance, and arises from a disk with an inner radius of $30~R_g$. This shows that the inferred relativistic smearing is highly sensitive to the assumed continuum shape.

Here we model only the first observation of this observing run (O1), and we set the spectral components in the timing section from fitting to the time-averaged spectral data alone. In future work, we will extend the modelling framework to fit to the later observations in this run (O2, O3, O4 of DM17, with ObsIDs 0760646301, 0760646401, 0760646501 respectively), where the source has declined further into the hard state. To this end we aim to develop a fully integrated fitting procedure, in which the spectral data is fit \textit{iteratively} with the cross-spectrum, in order to close the loop when determining the shape of the underlying spectral components of the timing model.

\section{Conclusions}
\label{sec:Conclusions}
In MD18a/b we developed a model for the spectral-timing properties of accreting hard state black hole binaries, whereby fluctuations in mass accretion rate were generated in - and propagated through - a spectrally stratified Comptonising hot flow. Here we have included key new features in this model to account for thermal reprocessing and reflection of the Comptonized X-rays illuminating the thin disc. We have fit this fully analytic spectral-timing model to the spectra, power spectra and lag-frequency spectra of key GX 339-4 XMM/NuSTAR observational data, which contain the strongest signal of thermal reverberation in X-ray binaries yet found. We fit the brightest hard state seen on the slow decline of the outburst in order to maximise signal-to-noise, while avoiding the source complexity seen in intermediate and fast rise hard states (see e.g. \citealt{G14}).

We model the spectrum with a thermal disc, two Comptonisation continua, and their reflections from that disc. The spectral model gives the contribution of each component in each energy band, used to develop the variability-emissivity model which is jointly fit to the power spectra in three different energy bands and the lag-frequency spectra between these bands (see also \citealt{R17b}, \citealt{V18}, MD18a/b). The variability and emission can be described as being self-similar throughout the hot flow (i.e. a constant generated fractional variability everywhere in the flow), with enhanced turbulence and emission only in a narrow ($1-3~R_g$) radial region where the thin disc and hot flow interact, centered at $19.2$~$R_g$.

These results support a truncated disc scenario in the hard state. They require a spectrum comprised of multiple Compton components in order to reproduce the propagation lags, and so they unambiguously motivate more complex spectral modeling than a single Comptonisation component and its reflection from the disc. These more complex spectral models reduce the relativistic smearing required in order to fit the iron line region, and also remove the requirement for highly super-solar iron abundances found in single-continuum fits. Our model of a disc truncated at $\sim 19 R_g$ in these hard state data can reproduce the time-averaged spectrum, the power spectra in different energy bands, the lags between these energy bands, and the lag-energy spectra including the reverberation signal from the disc. The untruncated disc models cannot explain the presence of propagation lags at energies above where the disc contributes to the emission, and predict a much shorter reverberation lag. In future work we will extend the technique developed here to explore the evolution of the disc radius in the remaining data from GX339-4, both from the slow outburst decline and on the fast rise.	

\section*{Acknowledgements}
We thank the anonymous referee for the highly insightful comments which helped to significantly improve the manuscript. RDM acknowledges the support of a Science and Technology Facilities Council (STFC) studentship through grant ST/N50404X/1. CD acknowledges the STFC through grant ST/P000541/1 for support. BDM acknowledges support from the Polish National Science Center grant Polonez 2016/21/P/ST9/04025. This work used the DiRAC Data Centric system at Durham University, operated by the Institute for Computational Cosmology on behalf of the STFC DiRAC HPC Facility (www.dirac.ac.uk). This equipment was funded by BIS National E-infrastructure capital grant ST/K00042X/1, STFC capital grant ST/H008519/1, and STFC DiRAC Operations grant ST/K003267/1 and Durham University. DiRAC is part of the National E-Infrastructure. This research has made use of data obtained through the High Energy Astrophysics Science Archive Research Center Online Service, provided by the NASA/Goddard Space Flight Center.

%%%%%%%%%%%%%%%%%%%% REFERENCES %%%%%%%%%%%%%%%%%%

%%%%%%%%%%%%%%%%% APPENDICES %%%%%%%%%%%%%%%%%%%%%
\appendix

\section{The Effect of Damping on the $\sigma_{rms}$-flux Relation}
\label{rms_flux_with_frequency}

An important and subtle potential consequence of the damping term which was not explored in MD18b was its effect on the root-mean-square-variability-flux relatiom ($\sigma_{rms}$-flux; \citealt{UMV05}). In real terms, the linear $\sigma_{rms}$-flux relation means that the absolute amplitude of rms variability increases linearly with the mean flux level, and this process is ubiquitous to X-ray signals from both BHXRBs and AGN. However, damping of propagated fluctuations can, in principle, introduce a frequency dependence to the slope of this relation. This dependence can arise due to the following process. A slow fluctuation first propagates from the variable disc inward, to modulate the variability generated in the \textit{soft} Compton region. Upon passing into the \textit{hard} Compton region, both the initial, long timescale fluctuation and the intermediate variability are damped. The fast variability generated in the \textit{hard} region is therefore modulated by only a damped form of the initial fluctuation, even though it preserves its variance. In contrast, the flux and $\sigma_{rms}$ associated with the intermediate variability are both damped by the same factor. This should result in a steeper $\sigma_{rms}$-flux relation for the intermediate (damped) variability than for the fast (undamped) variability. If too significant, this could result in a flux-dependent power spectral shape from a single spectral component. \cite{HVU12} show that on long timescales ($\sim$ 100~s), the power spectral shape for accreting BHXRBs is typically independent of flux within a fixed high energy band (2-13~keV). However it is unclear whether damping of the 1-5~s timescale modulations found here would be inconsistent with those longer-timescale measurements. For a direct check on this effect, we now compare the frequency-dependence of the $\sigma_{rms}$-flux relations for our model and data.

\begin{figure}
	\includegraphics[width=\columnwidth]{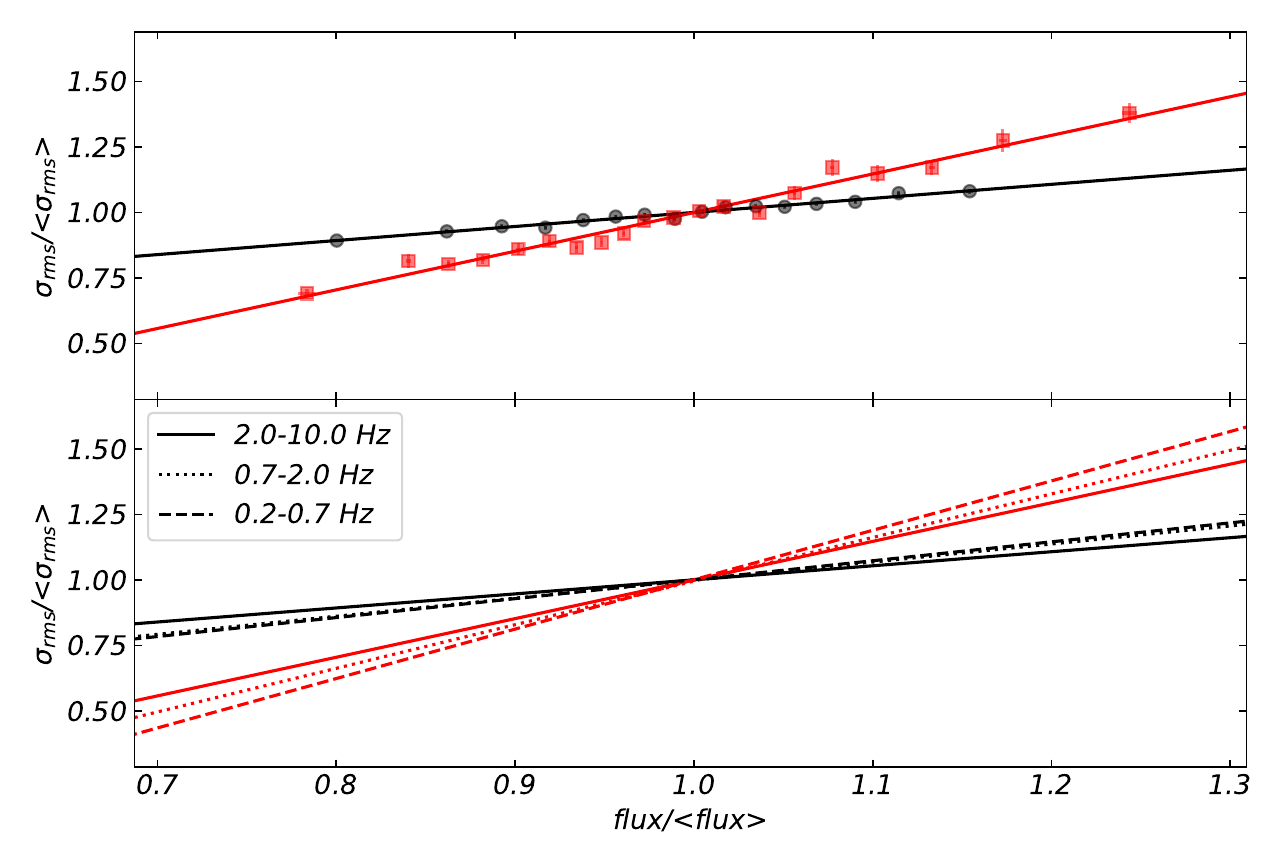}
	\caption{Panel (a): Flux-binned $\sigma_{rms}$-flux relations for the high-band (3-5~keV) data (black circles) and model (red squares) in the $2-10$~Hz frequency interval, with best fit lines. Panel (b): Best fit lines for $\sigma_{rms}$-flux relations for the data (black) and model (red) measured in three distinct frequency intervals indicated by different line styles. For all rms and flux values we have normalized by the respective mean values, in order to remove the trivial effect of the differential fractional rms in each frequency range. Ultimately we see a similar frequency-dependence to the $\sigma_{rms}$-flux relation gradient in both the data and the model.}
	\label{fig:rmsflux}
\end{figure}

In order to measure the $\sigma_{rms}$-flux relations for the model, we perform numerical simulations of the hot flow using the formalism of \cite{AU06}, updated from MD18a, with the parameters we have found for the flow here. We simulate light curves only in the high energy band (3-5~keV), as this most strongly samples the inner region, where all of the propagated variability should be found and the described frequency-dependence should be most pronounced. We then compute the $\sigma_{rms}$-flux relations for both the data and simulated light curves, using the power spectral method of \cite{UMV05} with 10-second segment lengths, integrated over three distinct frequency ranges ($0.2-7$~Hz, $0.7-2$~Hz, $2-10$~Hz). The result of this is shown in Fig.~\ref{fig:rmsflux}, where have normalized the rms and flux by their respective mean values.

For both the data and the model, we find dependence of the $\sigma_{rms}$-flux gradient upon the measured frequency interval. This dependence is not distinctly different between the simulated and observed light curves. The gradient of the lowest-frequency $\sigma_{rms}$-flux relation in the data is a factor 1.34 larger than the highest-frequency case. For the model, the lowest-frequency case is 1.32 times steeper than the highest-frequency case. The model is therefore comparable to the data in this regard. A similar frequency-dependence to the $\sigma_{rms}$-flux relation gradient is also seen in e.g. the 1996 hard state observations of Cygnus X-1 (\citealt{UMV05}) where the lower frequency variability also produces a steeper relation than the high-frequency variability. It is possible that the $\sigma_{rms}$-flux gradient in general \textit{is} sensitive to damping of the $1-5$~second modulations generated e.g. near the turbulent disc region or the outer flow, but that this effect becomes negligible compared to the longer timescale, larger amplitude modulations which change the flow-averaged mass accretion rate on $\sim$100~second timescales as seen in \cite{HVU12}. Indeed how the damping factors themselves would evolve between observations with very different flux values is also unclear; they are very likely coupled to the flow surface density, such that they change with changing mean mass accretion rate. Ultimately however, further work is required to determine whether the short-timescale frequency-dependence to the $\sigma_{rms}$-flux gradient we have seen here is ubiquitous.

\section{Spectral Comparison to a Single-Compton-Component Model}
\label{Single_vs_double_Compton}
\begin{figure*}
	\includegraphics[width=\textwidth]{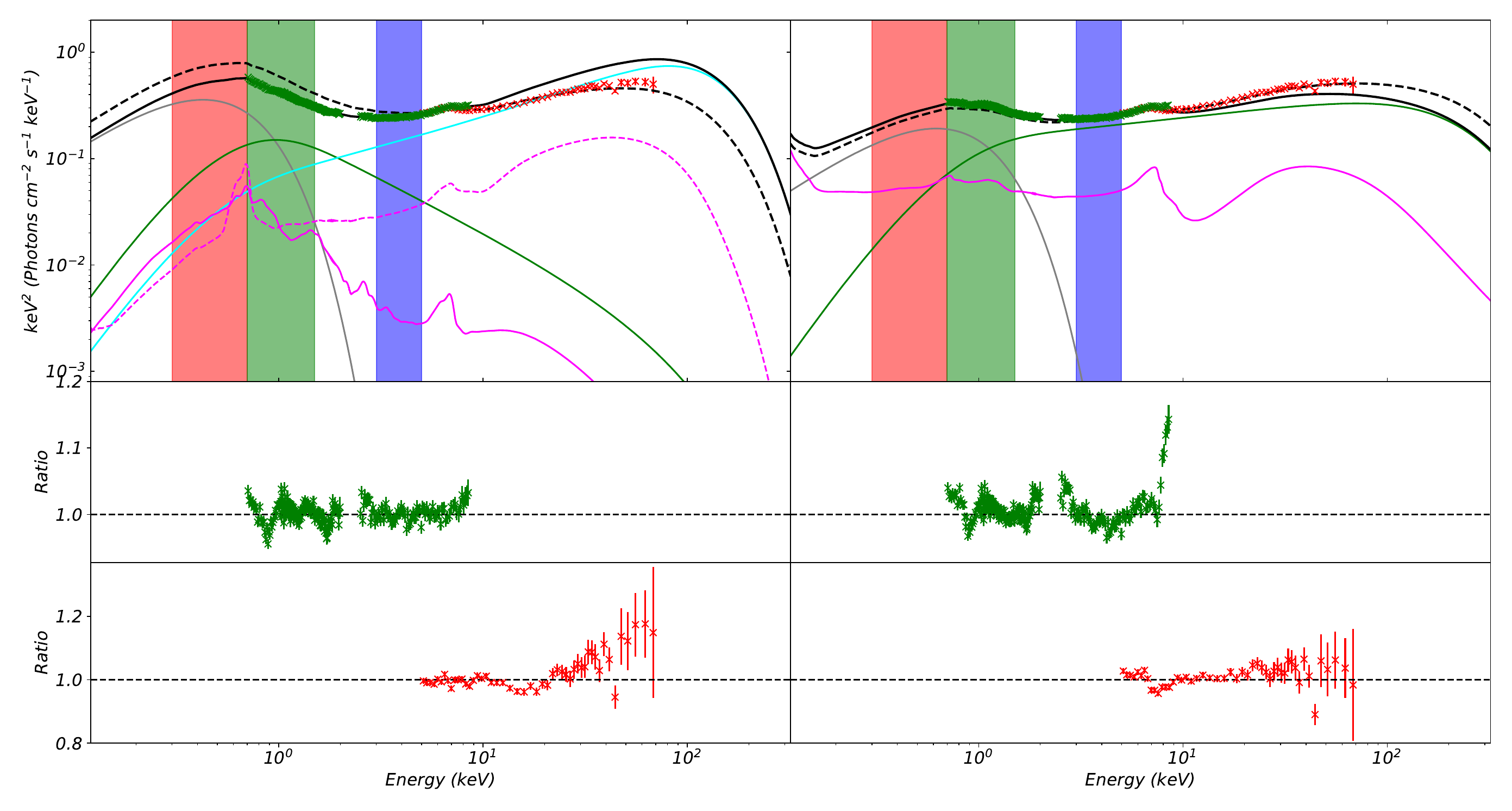}
	\caption{Left column: Spectral fit with two Compton components, identical to Fig.~\ref{fig:SED}. Colours also as in Fig.~\ref{fig:SED}. Green residuals are those for XMM data, red for NuSTAR. Right column: Spectral fit using a disc ({\tt{diskbb}}; solid grey line), a single Comptonisation component ({\tt{nthcomp}}; solid green line), and its disc reflection (solid magenta line), with a resultant model of {\tt{tbnew * (diskbb + nthcomp + kdblur * xillver * nthcomp)}}.}
	\label{fig:SED_both}
\end{figure*}
As we describe in section~\ref{sec:Discussion}, the same NuSTAR dataset as we have used here is also fit by \cite{WJ18}, using a model featuring only a single Comptonisation component rather than two. In the right-hand column of Fig.~\ref{fig:SED_both} we fit the O1 spectrum with a disc ({\tt{diskbb}}), a single Comptonisation component ({\tt{nthcomp}}), and the reflection of this Compton component from the disc, with parameters shown in Table~\ref{tab:SPECTRAL_one_comp}, in a manner similar to \cite{WJ18}. By setting the disc truncation to be small in this fit, the shape of the reflection spectrum around the FeK$\alpha$ line becomes highly peaked, such that the iron line can be reproduced. However this should also have the effect of producing significant line features below $1$~keV, which are not seen. To compensate for this, the single-Component model then requires an extremely super-solar iron abundance to smooth out the reflected flux below $1$~keV. Recent work by Tomsick et al. (2018; {\tt{reflionx}}) indicates that similar fits can be achieved with solar iron abundance by allowing the electron density to be much higher than standard (such that $n_e \approx 10^{20}$~$cm^{-3}$). Very recent fits of this type have lead to larger resultant truncation radii even in single-Compton-component models (\citealt{J19}). However the atomic physics relevant to electron densities this high are as yet unknown, making these calculations more tenuous, while there remains no mechanism in this picture to explain the observed hard lags.

On the other hand in the two-Compton-component model, the larger truncation radius gives rise to a smoother FeK$\alpha$ line profile. However given that this also produces less $<1$~keV reflected emission, the reflection normalization can be higher, with the \textit{soft} Compton component instead producing most of the flux in the $0.5-2$~keV range. In this way, the two-Compton-component model does not require an extreme iron abundance or an extreme truncation to reproduce the spectrum equally well (or in this case, even better; $\chi^2_{\nu,\,2\,comp} = 1.17$ vs $\chi^2_{\nu,\,1\,comp} = 1.36$). 

\begin{table}
	\begin{tabular}{ccc}
		\hline
		 				& XMM 					&NUSTAR\\
		\hline
		\vspace{+3pt}
		$N_H$ 				& $0.41^{+0.02}_{-0.01}$ 		& ==\\
		\vspace{+3pt}
		$kT_{in}$ 			& $0.262^{+0.010}_{-0.004}$		& ==\\
		\vspace{+3pt}
		$n_D$ 				& $6.02 \pm 0.04 \times 10^{3}$		& ==\\
		\vspace{+3pt}
		$\Gamma_S$ 			& $1.80^{+0.02}_{-0.03}$		& $1.73 \pm 0.03$\\
		\vspace{+3pt}
		$kT_{e,\,S}$ 			& $179^{+32}_{-15}$			& ==\\
		\vspace{+3pt}
		$n_S$ 				& $0.112 ^{+0.005}_{-0.004}$		&==\\
		\vspace{+3pt}
		$R_{in}$ 			& $2.69^{+0.11}_{-0.07}$ 		& ==\\
		\vspace{+3pt}
		$Incl$ 				& $48.4 \pm 0.2$			& ==\\
		\vspace{+3pt}
		$A_{Fe}$ 			& $8.67^{+0.33}_{-0.41}$ 		& ==\\
		\vspace{+3pt}
		log($\xi_i$) 			& $3.28^{+0.02}_{-0.03}$		& ==\\
		\hline
		$\chi^2/dof$ 			& \multicolumn{2}{c}{$3367/2467$ (combined)} \\	
		\hline
	\end{tabular}
	\caption[caption]{Parameter results of spectral fitting to O1 using the model {\tt{tbnew * (diskbb + nthcomp + kdblur * xillver * nthcomp)}}, fit simultaneously to XMM and NuSTAR data. Values not shown in the table are left to default.}
	\label{tab:SPECTRAL_one_comp}
\end{table}

\onecolumn
\section{An Updated Timing Formalism Including Disc Reprocessing and Reverberation}
\label{Timing Formalism}

In this appendix we describe how the timing formalism laid out in MD18b has been adapted to include thermal reverberation and reprocessing. For completeness we reiterate the steps of MD18b, highlighting where key changes have been made.

At each annulus, our model first calculates the generated mass accretion rate fluctuation profile in the frequency domain at each radius.
\vspace*{-3 pt}
\begin{equation}
\label{eq:Lorentzian}
|\tilde{\dot{m}}(r_n, f)|^2 \propto \frac{F_{var}(r_n)^2}{1+[f/f_{visc}(r_n)]^2}.
\end{equation} 
The $F_{var}(r)$ profile of equation equation~\ref{eq:newFvar} is incorporated into the model through this expression. This is also one of the two points at which the viscous frequency profile of equation~(\ref{eq:2visc}) influences the model, the other being the viscous travel time.

Once $\dot{m}$ profiles are generated, the $\dot{m}$ profile at each annulus can be propagated into the next in the Fourier domain through a convolution, following \cite{IvdK13}:
\begin{equation}
\label{eq:Mdot_propagation}
|\tilde{\dot{M}}(r_n, f)|^2 = |\tilde{\dot{m}}(r_n, f)|^2 \circledast |e^{2\pi i \Delta t_{(n-1)n} f} \tilde{\dot{M}}(r_{n-1}, f)|^2.
\end{equation}
Here the lag time is calculated as
\begin{equation}
\Delta t_{ln} = \sum_{k = l}^{n-1} d t_k = \sum_{k = l}^{n-1}\frac{dr_k}{r_k} t_{visc}(r_k) = dlog(r_k) \sum_{k = l}^{n-1} t_{visc}(r_k),
\end{equation}
where $t_{visc}(r_k) = 1/f_{visc}(r_k)$, this being the second point at which the $f_{visc}(r)$ profile influences our model output.

In equation~\ref{eq:Mdot_propagation} we have used a Green's response function to describe propagation from one annulus to the next, lagging on the local viscous timescale ($e^{2\pi i \Delta t_{ln} f}$) and damping fluctuations between spectral regions ($1/D_{ln}$, see equation~\ref{eq:DAMPING}). As described in Section~\ref{sec:Damping}, we have omitted the smoothing term used in MD18b. The Green's function used in equation~(\ref{eq:Mdot_propagation}) is therefore
\begin{equation}
\label{eq:GreensFunction}
\tilde{G}(r_l, r_n, f) =
\frac{1}{D_{ln}} e^{2\pi i \Delta t_{ln} f}.
\end{equation}
We can now calculate the exact propagated mass accretion rate profile at each annulus as
\begin{equation}
|\tilde{\dot{M}}(r_n, f)|^2 = \coprod^n_{l=1} \left|\frac{\tilde{\dot{m}}(r_l,f)}{D_{ln}}\right|^2,
\end{equation}
where $\coprod$ denotes a series of convolutions.
These mass accretion rate profiles can be converted to counts in a given energy band, $i$, using
the emissivity prescription and SED decomposition described in Section
\ref{sec:CorrelatedTurbulenceandEmissivity}. For the direct emission (i.e. not reflected or reprocessed), this effectively weights the propagated mass accretion
rate from the $n^{th}$ annulus by a factor, $w_{n,\,i,\,dir}$, given by
\begin{equation}
\label{eq:weights}
w_{n,\,i,\,dir} = \frac{\epsilon(r_n) r_n dr_n}{\sum\limits_{region} {\epsilon(r_n) r_n dr_n}}\int\displaylimits_{E = E_{i}^{min}}^{E_{i}^{max}} \bar{F}_{dir}(E, r_n)A_{eff}(E)e^{-N_H(E)\sigma_T} dE,
\end{equation}
where $A_{eff}(E)$ is the detector effective area, $N_H(E)$ is the galactic column absorption and $\sigma_T$ is the Thompson cross-section. The count rate of the direct emission for that band can then be written
\begin{equation}
C_{i,\,dir}(t) = \sum_{n = 1} ^{N} w_{n,\,i,\,dir} \dot{M}(r_n,t).
\end{equation}
Since the mean count rate of $\dot{M}(r_n, t)$ is normalized to $\dot{M}_0$, the mean count rate from the direct emission in band $i$ is then
\begin{equation}
\mu_{i,\,dir} = \sum_{n = 1}^{N} w_{n,\,i,\,dir} \dot{M}_0
\end{equation}
At this point we diverge from the formalism of MD18b by introducing a separate set of weights for the reflected/reprocessed emission, collectively the reverberated emission:
\begin{equation}
\label{eq:weights_rev}
w_{n,\,i,\,rev} = \frac{\epsilon(r_n) r_n dr_n}{\sum\limits_{region} {\epsilon(r_n) r_n dr_n}}\int\displaylimits_{E = E_{i}^{min}}^{E_{i}^{max}} \bar{F}_{rev}(E, r_n)A_{eff}(E)e^{-N_H(E)\sigma_T} dE.
\end{equation}
This results in a reverberated count rate in that band
\begin{equation}
C_{i,\,v}(t) = \sum_{n = 1} ^{N} w_{n,\,i,\,v} \dot{M}(r_n,t) \circledast IRF(t),
\end{equation}
where unlike the direct emission, we convolve this count rate with the impulse response function of Section~\ref{sec:Transferfn} to incorporate the reverberation delay. The IRF does not influence the light curve mean however, so similar to the direct light curve, the reverberation curve mean is 
\begin{equation}
\mu_{i,\,rev} = \sum_{n = 1}^{N} w_{n,\,i,\,rev} \dot{M}_0.
\end{equation}
Starting with the simple idea that a light curve is simply composed of direct and reverberated components originating from each annulus, we derive the overall power spectrum in the $i^{th}$ energy band to be
\begin{equation}
\begin{aligned}
\label{P1}
P_{i}(f) &\propto |\tilde{C}_{i,\,dir}(f)+\tilde{C}_{i,\,rev}(f)|^2\\
&\propto \sum_{l,\,n=1}^{N} [w_{l,\,i,\,dir}\tilde{\dot{M}}(r_l,f) + w_{l,\,i,\,rev}\tilde{\dot{M}}(r_l,f)TF(f)]^*[w_{n,\,i,\,dir}\tilde{\dot{M}}(r_n,f) + w_{n,\,i,\,rev}\tilde{\dot{M}}(r_n,f)TF(f)].
\end{aligned}
\end{equation}
In the case of unity mean mass accretion rate at each annulus, the cross-spectrum between annuli can be expressed,
\begin{equation}
\label{Mdotconv}
\tilde{\dot{M}}(r_l,f)^*\tilde{\dot{M}}(r_n,f) = \frac{e^{2 \pi i \Delta t_{ln}}}{D_{ln}} \left|\tilde{\dot{M}}(r_l,f)\right|^2.
\end{equation}
Combining equations~(\ref{P1})~\&~(\ref{Mdotconv}) and including the relevant normalization, it can then be shown that the power spectrum of a light curve in the $i^{th}$ energy band is exactly described by
\begin{equation}
\label{eq:PSDprop2band}
\begin{aligned}
P_{i}(f) = &\frac{2dt^2}{(\mu_{i,\,dir} + \mu_{i,\,rev} + L_{i,\,th,\,const})^2 T} \times \\
&\sum_{n=1}^{N}(w_{n,\,i,\,dir}^2 + 2w_{n,\,i,\,dir}w_{n,\,i,\,rev}TF_{re}(f) + w_{n,\,i,\,rev}^2|TF(f)|^2)|\tilde{\dot{M}}(r_n,f)|^2\\ 
&+ 2\sum_{l=1}^{n-1}\frac{|\tilde{\dot{M}}(r_l,f)|^2}{D_{ln}} \left\{ TF_{im}(f)\text{sin}(2 \pi \Delta t_{ln})( w_{l,\,i,\,dir}w_{n,\,i,\,rev}-w_{l,\,i,\,rev}w_{n,\,i,\,dir})\right.\\
& \left. + \text{cos}(2 \pi \Delta t_{ln} f)(w_{l,\,i,\,dir}w_{n,\,i,\,dir} + TF_{re}(f)(w_{l,\,i,\,dir}w_{n,\,i,\,rev}+w_{l,\,i,\,rev}w_{n,\,i,\,dir})+w_{l,\,i,\,rev}w_{n,\,i,\,rev}|TF(f)|^2) \right\}. 
\end{aligned}
\end{equation}
where $TF_{re}$ and $TF_{im}$ are the real and imaginary parts of the transfer function respectively, and $L_{i,\,th,\,const}$ is the luminosity of the constant component of the thermal disc in band $i$ (i.e. that left over after subtracting the intrinsically variable and reprocessed thermal luminosity). In the limit that the reverberation signal is zero (i.e. $w_{n,\,i,\,rev}=0$ for all $n,\,i$), this expression reduces to equation~(A12) of MD18b, which did not include the effects of reverberation. The same logic can be used to show that the real part of the cross spectrum between two bands $i$ and $j$ is
\begin{equation}
\label{eq:CrossRe}
\begin{aligned}
	\mathfrak{Re}[\Gamma_{ij}(f)]=&\frac{2dt^2}{(\mu_{i,\,dir} + \mu_{i,\,rev} + L_{i,\,th,\,const}) (\mu_{j,\,dir} + \mu_{j,\,rev} + L_{j,\,th,\,const}) T} \times\\
	&\sum_{n=1}^{N}(w_{n,\,i,\,dir}w_{n,\,j,\,dir} + (w_{n,\,i,\,dir}w_{n,\,j,\,rev} + w_{n,\,j,\,dir}w_{n,\,i,\,rev}) TF_{re}(f) + (w_{n,\,i,\,rev}w_{n,\,j,\,rev})|TF(f)|^2)|\tilde{\dot{M}}(r_n,f)|^2\\
	&+ \sum_{l=1}^{n-1} \frac{|\tilde{\dot{M}}(r_l,f)|^2}{ D_{ln}}\left[ \text{cos}(2 \pi \Delta t_{ln} f) \right.
		\!\begin{aligned}[t]
		&\left(w_{l,\,i,\,dir}w_{n,\,j,\,dir} + w_{n,\,i,\,dir}w_{l,\,j,\,dir}\right. \\
		& +TF_{re}(f)(w_{l,\,i,\,dir}w_{n,\,j,\,rev}+w_{l,\,i,\,rev}w_{n,\,j,\,dir} + w_{n,\,i,\,dir}w_{l,\,j,\,rev} + w_{n,\,i,\,rev}w_{l,\,j,\,dir})\\
		&\left.+(w_{l,\,i,\,rev}w_{n,\,j,\,rev} + w_{n,\,i,\,rev}w_{l,\,j,\,rev})|TF(f)|^2) \right) \\
		\end{aligned} \\
	& \left. - TF_{im}(f)\text{sin}(2 \pi \Delta t_{ln})(w_{l,\,i,\,dir}w_{n,\,j,\,rev} - w_{l,\,i,\,rev}w_{n,\,j,\,dir} - w_{n,\,i,\,dir}w_{l,\,j,\,rev} + w_{n,\,i,\,rev}w_{l,\,j,\,dir})\right],
\end{aligned}
\vspace{-10pt}
\end{equation}
and the imaginary part is
\begin{equation}
\label{eq:CrossIm}
\begin{aligned}
	\mathfrak{Im}[\Gamma_{ij}(f)] =&\frac{2dt^2}{(\mu_{i,\,dir} + \mu_{i,\,rev} + L_{i,\,th,\,const}) (\mu_{j,\,dir} + \mu_{j,\,rev} + L_{j,\,th,\,const}) T} \times\\
	&\sum_{n=1}^{N}(w_{n,\,i,\,dir}w_{n,\,j,\,rev} - w_{n,\,j,\,dir}w_{n,\,i,\,rev}) TF_{im}(f)|\tilde{\dot{M}}(r_n,f)|^2\\
	&+ \sum_{l=1}^{n-1}\frac{|\tilde{\dot{M}}(r_l,f)|^2}{ D_{ln}}[ \text{sin}(2 \pi \Delta t_{ln} f)
		\!\begin{aligned}[t]
		&\left(w_{l,\,i,\,dir}w_{n,\,j,\,dir} - w_{n,\,i,\,dir}w_{l,\,j,\,dir} \right. \\
		& + TF_{re}(f)(w_{l,\,i,\,dir}w_{n,\,j,\,rev}+w_{l,\,i,\,rev}w_{n,\,j,\,dir} - w_{n,\,i,\,dir}w_{l,\,j,\,rev} - w_{n,\,i,\,rev}w_{l,\,j,\,dir})\\
		&\left.+(w_{l,\,i,\,rev}w_{n,\,j,\,rev} - w_{n,\,i,\,rev}w_{l,\,j,\,rev})|TF(f)|^2) \right) \\
		\end{aligned}\\
	& \left. + TF_{im}(f)\text{cos}(2 \pi \Delta t_{ln})(w_{l,\,i,\,dir}w_{n,\,j,\,ref} - w_{l,\,i,\,rev}w_{n,\,j,\,dir} + w_{n,\,i,\,dir}w_{l,\,j,\,rev} - w_{n,\,i,\,rev}w_{l,\,j,\,dir})\right]. 
\end{aligned}
\end{equation}
The frequency-resolved time lag between bands $i$ and $j$ is then
\begin{equation}
\label{eq:tau_analytic}
tan(2\pi f \tau_{ij})=\frac{\mathfrak{Im}[\Gamma_{ij}(f)]}{\mathfrak{Re}[\Gamma_{ij}(f)]}.
\end{equation}

\bsp
\label{lastpage}

\begin{thebibliography}{99}
	
\bibitem[\protect\citeauthoryear{Ar\'{e}valo \& Uttley}{2006}]{AU06}
Ar\'{e}valo P., Uttley P., 2006, MNRAS, 367, 801 

\bibitem[\protect\citeauthoryear{Arnaud, Borkowski \& Harrington}{1996}]{ABH96}
Arnaud K., Borkowski K.J., Harrington J.P., 1996, ApJ, 462, L75

%\bibitem[\protect\citeauthoryear{Axelsson et al.}{2005}]{A05}
%Axelsson M., Borgonovo L., Larsson S., 2005, A\&A, 438, 999

%\bibitem[\protect\citeauthoryear{Axelsson et al.}{2008}]{A08}
%Axelsson M., Hjalmarsdotter L., Borgonovo L., Larsson S., 2008, A\&A, 490, 253

\bibitem[\protect\citeauthoryear{Axelsson \& Done}{2018}]{AD18}
Axelsson M., Done C., 2018, MNRAS, 480 (1), 751

\bibitem[\protect\citeauthoryear{Balbus \& Hawley}{1998}]{BH98}
Balbus S.A., Hawley J.F., 1998, RvMP, 70, 1

\bibitem[\protect\citeauthoryear{Basak \& Zdziarski}{2016}]{BZ16}
Basak R., Zdziarski A.A., 2016, MNRAS, 458, 2199

\bibitem[\protect\citeauthoryear{Basak et al.}{2017}]{B17}
Basak R., Zdziarski A.A., Parker M., Islam N., 2017, MNRAS, 472, 4220

%\bibitem[\protect\citeauthoryear{Belloni et al.}{2002}]{B02}
%Belloni T., Psaltis D., van der Klis M., 2002, ApJ, 572, 392

\bibitem[\protect\citeauthoryear{Belloni et al.}{2005}]{B05}
Belloni T., Homan J., Casella P. et al., 2002, A\&A, 440, 207

%\bibitem[\protect\citeauthoryear{Beloborodov}{1999}]{B99}
%Beloborodov A.M., 1999, ApJ, 510, L123

%\bibitem[\protect\citeauthoryear{Blaes}{2013}]{B13}
%Blaes O., 2013, Space Sci. Rev., 103, 21

%\bibitem[\protect\citeauthoryear{Blandford \& McKee}{1982}]{BM82}
%Blandford R.D., McKee C.F., 1982, ApJ, 255, 419

\bibitem[\protect\citeauthoryear{Chainakun \& Young}{2012}]{CY12}
Chainakun P., Young A.J., 2012, MNRAS, 420 (2), 1145

%\bibitem[\protect\citeauthoryear{Churazov, Gilfanov \& Revnivtsev}{2001}]{CGR01}
%Churazov E., Gilfanov M., Revnivtsev M., 2001, MNRAS, 321, 759

%\bibitem[\protect\citeauthoryear{Di Salvo et al.}{2001}]{DS01}
%Di Salvo T., Done C., $\dot{Z}$ycki P.T., Burderi L., Robba N.R., 2001, ApJ, 547, 1024

\bibitem[\protect\citeauthoryear{De Marco et al.}{2015}]{DM15}
De Marco B., Ponti G., Mu\~{n}oz-Darias T., Nandra K., 2015, ApJ, 814, 50

\bibitem[\protect\citeauthoryear{De Marco \& Ponti}{2016}]{DM16}
De Marco B., Ponti G., 2016, ApJ, 826, 70

\bibitem[\protect\citeauthoryear{De Marco et al.}{2017}]{DM17}
De Marco B., Ponti G., Petrucci P.O. et al., 2017, MNRAS, 471, 1475 (DM17)

\bibitem[\protect\citeauthoryear{Done, Gierli\'{n}ski \& Kubota}{2007}]{DGK07}
Done C., Gierli\'{n}ski M., Kubota A., 2007, A\&ARv, 15, 1

\bibitem[\protect\citeauthoryear{Done \& Diaz-Trigo}{2010}]{DDT10}
Done C., Diaz-Trigo M., 2010, MNRAS, 407 (4), 2287

%\bibitem[\protect\citeauthoryear{Dunn et al.}{2010}]{D10}
%Dunn R.J.H, Fender R.P., Kording E.G., Belloni T., Cabanac C., 2010, MNRAS, 403, 61

\bibitem[\protect\citeauthoryear{Dzie\l{}ak}{2018}]{DAM18}
Dzie\l{}ak M.C., Zdziarski A.A., Szanecki M., De Marco B., Nied\'{z}wiecki A., Markowitz A., 2018, MNRAS, 485 (3), 3845

\bibitem[\protect\citeauthoryear{Esin, McClintock \& Narayan}{1997}]{EMN97}
Esin A.A., McClintock J.E., Narayan R., 1997, ApJ, 489 (2), 865

\bibitem[\protect\citeauthoryear{Fabian et al.}{2014}]{F14}
Fabian A.C., Parker M.L., Wilkins D.R. et al., 2014, MNRAS, 439, 2307

%\bibitem[\protect\citeauthoryear{Foreman-Mackey et al.}{2013}]{DFM13}
%Foreman-Mackey D., Hogg D.W., Lang D., Goodman J., 2013, Publ. Astron. Soc. Pac, 125, 306

\bibitem[\protect\citeauthoryear{Fragile et al.}{2007}]{F07}
Fragile P.C., Blaes O.M., Anninos P., Salmonson J.D., 2009, ApJ, 668, 417

\bibitem[\protect\citeauthoryear{Frank, King \& Raine}{2002}]{FKR}
Frank J., King A., Raine D., 2002, \textit{Accretion Power in Astrophysics: Third Edition}, Cambridge University Press, pp.~83-84

\bibitem[\protect\citeauthoryear{F{\"u}rst et al.}{2015}]{F15}
F{\"u}rst F., Nowak M.A., Tomsick J.A., 2015, ApJ, 808(2), 122

\bibitem[\protect\citeauthoryear{Garc\'{i}a et al.}{2015}]{G15}
Garc\'{i}a J.A., Steiner J.F, McClintock J.E., Remillard R.A., Grinberg V., Dauser T., 2015, ApJ, 813, 84

\bibitem[\protect\citeauthoryear{Gardner \& Done}{2014}]{GD14}
Gardner E., Done C., 2014, MNRAS, 442, 2456

\bibitem[\protect\citeauthoryear{Gardner \& Done}{2017}]{GD17}
Gardner E., Done C., 2017, MNRAS, 470, 3591

\bibitem[\protect\citeauthoryear{Gendreau, Arzoumanian \& Okajima}{2012}]{NICER12}
Gendreau K.C., Arzoumanian Z., Okajima T., 2012, Proc. SPIE 8443, Space Telescopes and Instrumentation 2012: Ultraviolet to Gamma Ray, 844313

%\bibitem[\protect\citeauthoryear{Generozov et al.}{2014}]{GBFH14}
%Generozov A., Blaes O., Fragile P.C., Henisey K.B., 2014, ApJ, 780, 81

%\bibitem[\protect\citeauthoryear{Gierli{\'n}ski et al.}{1997}]{G97}
%Gierli{\'n}ski M., Zdziarski A.A., Done C. et al., 1997, MNRAS, 288, 958

\bibitem[\protect\citeauthoryear{Gierli{\'n}ski \& Done}{2004}]{GD04}
Gierli{\'n}ski M., Done C., 2004, MNRAS, 347, 885

%\bibitem[\protect\citeauthoryear{Gierli{\'n}ski, Done \& Page}{2009}]{GDP09}
%Gierli{\'n}ski M., Done C., Page K., MNRAS, 392, 1106

%\bibitem[\protect\citeauthoryear{Gilfanov, Churazov \& Revnivtsev}{2000}]{GCR00}
%Gilfanov M., Churazov E., Revnivtsev M., 2000, MNRAS, 316, 923

\bibitem[\protect\citeauthoryear{Grinberg et al.}{2014}]{G14}
Grinberg V., Pottshmidt K., B{\"o}ck M. et al., 2014, A\&A, 565, A1

\bibitem[\protect\citeauthoryear{Heil, Vaughan \& Uttley}{2012}]{HVU12}
Heil L.M., Vaughan S. \& Uttley P., 2012, MNRAS, 422, 3620

%\bibitem[\protect\citeauthoryear{Haardt \& Maraschi}{1993}]{HM93}
%Haardt F., Maraschi L., 1993, ApJ, 413, 507

\bibitem[\protect\citeauthoryear{Heida et al.}{2017}]{HJTC17}
Heida M., Jonker P.G., Torres M.A.P., Chiavassa A., ApJ, 846 (2), 132

%\bibitem[\protect\citeauthoryear{Henisey, Blaes \& Fragile}{2012}]{HBF12}
%Henisey K.B., Blaes O.M., Fragile P.C., 2012, ApJ, 761, 18

%\bibitem[\protect\citeauthoryear{Hogg \& Reynolds}{2017}]{HR17}
%Hogg J.D., Reynolds C.S., 2017, ApJ, 834, 80

%\bibitem[\protect\citeauthoryear{Ibragimov et al.}{2005}]{I05}
%Ibragimov A., Poutanen J., Gilfanov M., Zdziarski A.A. \& Shrader C.R., 2005, MNRAS, 362, 1435

\bibitem[\protect\citeauthoryear{Ingram et al.}{2009}]{IDF09}
Ingram A., Done C., Fragile P.C., 2009, MNRAS, 397 (1), L101

\bibitem[\protect\citeauthoryear{Ingram \& Done}{2011}]{ID11}
Ingram A., Done C., 2011, MNRAS, 415 (3), 2323 (ID11)

\bibitem[\protect\citeauthoryear{Ingram \& Done}{2012a}]{ID12a}
Ingram A., Done C., 2012, MNRAS, 419, 2369

%\bibitem[\protect\citeauthoryear{Ingram \& Done}{2012b}]{ID12b}
%Ingram A., Done C., 2012, MNRAS, 427, 934

\bibitem[\protect\citeauthoryear{Ingram \& van der Klis}{2013}]{IvdK13}
Ingram A., van der Klis M., 2013, MNRAS, 434, 1476

\bibitem[\protect\citeauthoryear{Ingram et al.}{2016}]{I16}
Ingram A., van der Klis M., Middleton M. et al., 2016, MNRAS, 461 (2), 1967

\bibitem[\protect\citeauthoryear{Ingram et al.}{2017}]{I17}
Ingram A., van der Klis M., Middleton M., Altamirano D., Uttley P., 2017, MNRAS, 464 (3), 2979

\bibitem[\protect\citeauthoryear{Jiang et al.}{2019}]{J19}
Jiang J., Fabian A.C, Wang J. et al., 2019, MNRAS, \textit{in press}

\bibitem[\protect\citeauthoryear{Kara et al.}{2013}]{K13}
Kara E., Fabian A.C., Cackett E.M. et al., 2013, MNRAS, 428, 2795

\bibitem[\protect\citeauthoryear{Kara et al.}{2019}]{K19}
Kara E., Steiner J.F., Fabian A.C. et al., 2019, Nature, 565 (7738), 198

%\bibitem[\protect\citeauthoryear{Kawano et al.}{2017}]{K17}
%Kawano T., Done C., Yamada S., Takahashi H., Axelsson M., Fukuzawa Y., 2017, PASJ, 69(2), 36

%\bibitem[\protect\citeauthoryear{Klein-Wolt \& van der Klis}{2008}]{KWvdK08}
%Klein-Wolt M., van der Klis M., 2008, ApJ, 675, 1407

\bibitem[\protect\citeauthoryear{Kolehmainen, Done \& Diaz Trigo}{2014}]{KDD14}
Kolehmainen M., Done C., Diaz Trigo M., 2014, MNRAS, 437, 613

\bibitem[\protect\citeauthoryear{Kotov, Churazov \& Gilfanov}{2001}]{KCG01}
Kotov O., Churazov E., Gilfanov M., 2001, MNRAS, 327, 799

%\bibitem[\protect\citeauthoryear{Liska et al.}{2017}]{L17}
%Liska M., Hesp C., Tchekhovskoy A., Ingram A., van der Klis M., Markoff S., 2017, MNRAS, 474, L81

%\bibitem[\protect\citeauthoryear{Lubow, Ogilvie \& Pringle}{2002}]{LOP02}
%Lubow S.H., Ogilvie G.I., Pringle J.E., 2002, MNRAS, 337, 706

\bibitem[\protect\citeauthoryear{Lyubarskii}{1997}]{L97}
Lyubarskii Y.E., 1997, MNRAS, 292, 679

\bibitem[\protect\citeauthoryear{Mahmoud \& Done}{2018a}]{MD18a}
Mahmoud R.D., Done C., 2018, MNRAS, 473, 2084 (MD18a)

\bibitem[\protect\citeauthoryear{Mahmoud \& Done}{2018b}]{MD18b}
Mahmoud R.D., Done C., 2018, MNRAS, 480 (3), 4040 (MD18b)

\bibitem[\protect\citeauthoryear{Makishima et al.}{2008}]{M08}
Makishima K., Takahashi H., Yamada S. et al., 2008, PASJ, 60, 585

%\bibitem[\protect\citeauthoryear{Mastroserio, Ingram \& van der Klis}{2018}]{MIvdK18}
%Mastroserio G., Ingram A., van der Klis M., 2018, MNRAS, 475, 4027 

\bibitem[\protect\citeauthoryear{McClintock \& Remillard}{2006}]{MR06}
McClintock J.E., Remillard R.A., 2006, in \textit{Compact Stellar X-Ray Sources}, Ch. 4, Cambridge University Press, ed. Lewin, W.H.G. \& van der Klis, M.

%\bibitem[\protect\citeauthoryear{Markoff, Nowak \& Wilms}{2005}]{MNW05}
%Markoff S., Nowak M.A. \& Wilms J., 2005, ApJ, 635, 1203

\bibitem[\protect\citeauthoryear{Miller et al.}{2006}]{M06}
Miller J.M., Homan J., Steeghs D. et al., 2006, ApJ, 653, 525

%\bibitem[\protect\citeauthoryear{Misra et al.}{2017}]{M17}
%Misra R., Yadav J.S., Chauhan J.V. et al., 2017, ApJ, 835 (2), 195

\bibitem[\protect\citeauthoryear{Miyamoto \& Kitamoto}{1989}]{MK89}
Miyamoto A., Kitamoto S., 1989, Nature, 342, 773

\bibitem[\protect\citeauthoryear{Mu\~{n}oz-Darias, Motta \& Belloni}{2011}]{MMB11}
Mu\~{n}oz-Darias T., Motta S., Belloni T.M., 2011, MNRAS, 410, 679 

%\bibitem[\protect\citeauthoryear{Mushtukov, Ingram \& van der Klis}{2017}]{MIvdK17}
%Mushtukov A.A., Ingram A., van der Klis M., 2017, MNRAS, 474, 2259

%\bibitem[\protect\citeauthoryear{Narayan, Kato \& Honma}{1997}]{NKH97}
%Narayan R., Kato S., Honma F., 1997, ApJ, 476, 49

%\bibitem[\protect\citeauthoryear{Narayan \& Yi}{1995}]{NY95}
%Narayan R., Yi I., 1995, ApJ, 452, 710

%\bibitem[\protect\citeauthoryear{Noble \& Krolik}{2009}]{NK09}
%Noble S.C., Krolik J.H., 2009, ApJ, 703, 964

\bibitem[\protect\citeauthoryear{Nowak \text{et al.}}{1999}]{N99}
Nowak M.A., Vaughan B.A., Wilms J., Dove J.B., Begelman M.C., 1999, ApJ, 510, 874

%\bibitem[\protect\citeauthoryear{Nowak}{2000}]{N00}
%Nowak M.A., 2000, MNRAS, 318, 361

\bibitem[\protect\citeauthoryear{Nowak}{2011}]{N11}
Nowak M.A., Hanke M., Trowbridge S.N. et al., 2011, ApJ, 728, 13

\bibitem[\protect\citeauthoryear{Novikov \& Thorne}{1973}]{NT73}
Novikov I.D., Thorne K.S., 1973, blho.conf, 343

%\bibitem[\protect\citeauthoryear{Papadakis \& Lawrence}{1993}]{PL93}
%Papadakis I.E., Lawrence A., 1993, MNRAS, 261, 612

\bibitem[\protect\citeauthoryear{Parker at al.}{2016}]{P16}
Parker M.L., Tomsick J.A., Kennea J.A. et al., 2016, ApJ \textit{Letts.}, 821 (1), L6

%\bibitem[\protect\citeauthoryear{Plant at al.}{2015}]{P15}
%Plant D.S., Fender R.P., Ponti G., Mu\~{n}oz-Darias T., Coriat M., 2015, A&A, 573, A120

%\bibitem[\protect\citeauthoryear{Pottschmidt et al.}{2003}]{P03}
%Pottschmidt K., Wilms J., Nowak M.A. et al., 2003, A\&A, 407, 1039

%\bibitem[\protect\citeauthoryear{Poutanen \& Coppi}{1998}]{PC98}
%Poutanen J., Coppi P.S., 1998, Phys. Scripta, T77, 57

%\bibitem[\protect\citeauthoryear{Poutanen \& Vurm}{2009}]{PV09}
%Poutanen J., Vurm I., 2009, ApJ, 690, L97

\bibitem[\protect\citeauthoryear{Poutanen \& Veledina}{2014}]{PV14}
Poutanen J., Veledina A., 2014, Space Sci. Rev., 183, 61

\bibitem[\protect\citeauthoryear{Poutanen, Veledina \& Zdziarski}{2018}]{PVZ18}
Poutanen J., Veledina A., Zdziarski A.A., 2018, A\&A, 614, A79

%\bibitem[\protect\citeauthoryear{Rapisarda, Ingram \& van der Klis}{2014}]{RIvdK14}
%Rapisarda S., Ingram A., van der Klis M., 2014, MNRAS, 440, 2882

%\bibitem[\protect\citeauthoryear{Rapisarda et al.}{2016}]{R16}
%Rapisarda S., Ingram A., Kalamkar M., van der Klis M., 2016, MNRAS, 462, 4078

\bibitem[\protect\citeauthoryear{Rapisarda, Ingram \& van der Klis}{2017a}]{R17a}
Rapisarda S., Ingram A., van der Klis M., 2017, MNRAS, 469 (2), 2017

\bibitem[\protect\citeauthoryear{Rapisarda, Ingram \& van der Klis}{2017b}]{R17b}
Rapisarda S., Ingram A., van der Klis M., 2017, 472, 3821

%\bibitem[\protect\citeauthoryear{Reis, Fabian \& Miller}{2010}]{RFM10}
%Reis R.C, Fabian A.C, Miller J.M., 2010, MNRAS, 402, 836

\bibitem[\protect\citeauthoryear{Remillard \& McClintock}{2006}]{RM06}
Remillard R.A., McClintock J.E., 2006, ARA\&A, 44 (1), 49

\bibitem[\protect\citeauthoryear{Revnivtsev, Gilfanov \& Churazov}{1999}]{RGC99}
Revnivtsev M., Gilfanov M. \& Churazov E., 1999, A\&A, 347, L23

%\bibitem[\protect\citeauthoryear{Revnivtsev et al.}{2011}]{R11}
%Revnivtsev M., Potter S., Kniazev A., Burenin R., Buckley D.A.H. \& Churazov E., 2011, MNRAS, 411, 1317

%\bibitem[\protect\citeauthoryear{Rykoff et al.}{2007}]{Ry07}
%Rykoff, E.S., Miller J.M, Steeghs D., Torres M.A.P., 2007, ApJ, 666, 1129

\bibitem[\protect\citeauthoryear{Shakura \& Sunyaev}{1973}]{SS73}
Shakura N.I., Sunyaev R.A., 1973, A\&A, 24, 337

\bibitem[\protect\citeauthoryear{Steiner et al.}{2010}]{S10}
Steiner J.F., McClintock J.E., Remillard R.A., Gou L., Yamada S., Narayan R., 2010, ApJ, 718, L117

%\bibitem[\protect\citeauthoryear{Tomsick et al.}{2008}]{T08}
%Tomsick J.A., Kalemci E., Kaaret P. et al., 2008, ApJ, 680, 593

%\bibitem[\protect\citeauthoryear{Tomsick et al.}{2014}]{T14}
%Tomsick J.A. et al., 2014, ApJ, 780, 78

\bibitem[\protect\citeauthoryear{Tomsick et al.}{2018}]{TPG18}
Tomsick J.A., Parker M.L., Garc\'{i}a J.A., 2018, ApJ, 855, 3

%\bibitem[\protect\citeauthoryear{Timmer \& K{\"o}nig}{1995}]{TK95}
%Timmer J., K{\"o}nig M., 1995, A\&A, 300, 707

%\bibitem[\protect\citeauthoryear{Torii et al.}{2011}]{T11}
%Torii S., Yamada S., Makishima K, et al., 2011, PASJ, 63, S771


\bibitem[\protect\citeauthoryear{Uttley, M$^c$Hardy \& Vaughan}{2005}]{UMV05}
Uttley P., M$^c$Hardy I.M., Vaughan S., 2005, MNRAS, 359, 346

\bibitem[\protect\citeauthoryear{Uttley et al.}{2011}]{U11}
Uttley P., Wilkinson T., Cassatella P., Wilms E., Pottschmidt K., Hanke M., B{\"o}ck M., 2011, MNRAS, 414, L60

\bibitem[\protect\citeauthoryear{Uttley et al.}{2014}]{U14}
Uttley P., Cackett E.M., Fabian A.C., Kara E., Wilkins D.R., 2014, Astron. Astrophys. Rev., 22, 72

\bibitem[\protect\citeauthoryear{van der Klis}{1989}]{vdK89}
van der Klis M., 1989, in \textit{Timing Neutron Stars: proceedings of the NATO Advanced Study Institute on Timing Neutron Stars}, p.27, Kluwer Academic / Plenum Publishers, New York, ed. {\"O}gelman H. \& van den Heuvel E.P.J.

\bibitem[\protect\citeauthoryear{Vaughan \& Nowak}{1997}]{VN97}
Vaughan B.A., Nowak M.A., 1997, ApJ, 474, L43

\bibitem[\protect\citeauthoryear{Veledina}{2016}]{V16}
Veledina A., 2016, ApJ, 832, 181

\bibitem[\protect\citeauthoryear{Veledina}{2018}]{V18}
Veledina A., 2018, MNRAS, 481, 4236

\bibitem[\protect\citeauthoryear{Wang-Ji et al.}{2018}]{WJ18}
Wang-Ji J., Garc\'{i}a J.A., Steiner J.F. et al., 2018, ApJ, 855 (1), 61

\bibitem[\protect\citeauthoryear{Welsh \& Horne}{1991}]{WH91}
Welsh W.F., Horne K., 2016, ApJ, 379, 586

\bibitem[\protect\citeauthoryear{Wijnands \& van der Klis}{1999}]{WvdK99}
Wijnands R., van der Klis M., 1999, ApJ, 522 (2), 965

\bibitem[\protect\citeauthoryear{Wilkinson \& Uttley}{2009}]{WU09}
Wilkinson T., Uttley P., 2009, MNRAS, 397, 666

\bibitem[\protect\citeauthoryear{Wilms, Allen \& McCray}{2000}]{WAM00}
Wilms J., Allen A., McCray R. 2000, ApJ, 542, 914

%\bibitem[\protect\citeauthoryear{Yamada et al.}{2013}]{Y13}
%Yamada S., Makishima K., Done C., Torii S., Noda H., Sakurai S., 2013, PASJ, 65, 80

\bibitem[\protect\citeauthoryear{Yuan, Quataert \& Narayan}{2003}]{YQN03}
Yuan F., Quataert E. \& Narayan R., 2003, ApJ, 598 (1), 301

\bibitem[\protect\citeauthoryear{Zdziarski, Johnson \& Magdziarz}{1996}]{ZJM96}
Zdziarski A.A., Johnson W.N., Magdziarz P., 1996, MNRAS, 283, 193

\end{thebibliography}
\end{document}